\documentclass[%
aps,%
 amsmath,amssymb,
 reprint,%
 superscriptaddress,
nofootinbib,
floatfix
]{revtex4-1}

\usepackage{graphicx}
\usepackage{dcolumn}
\usepackage{bm}
\usepackage{xcolor}
\graphicspath{{Images/}}

\usepackage[caption=false]{subfig}
\usepackage[normalem]{ulem}
\usepackage{braket}
\usepackage{esdiff}
\usepackage{hyperref}
\usepackage{appendix}
\usepackage{mathtools}
\usepackage{amsmath}
\usepackage{calc}
\usepackage{booktabs}

\begin{document}

\preprint{AAPM/123-QED}

\title{Entanglement and collective flavor oscillations in a dense neutrino gas}

\author{Michael J.\ Cervia}
\email{cervia@wisc.edu}
\affiliation{%
Department of Physics, University of Wisconsin--Madison,
Madison, Wisconsin 53706, USA
}%
\author{Amol V.\ Patwardhan}
\email{apatwardhan2@berkeley.edu} 
\affiliation{%
Department of Physics, University of Wisconsin--Madison,
Madison, Wisconsin 53706, USA
}%
\affiliation{
Department of Physics, University of California, Berkeley, CA 94720-6300, USA} 
\author{\\ A.\ B.\ Balantekin}
\email{baha@physics.wisc.edu}
\affiliation{%
Department of Physics, University of Wisconsin--Madison,
Madison, Wisconsin 53706, USA
}%
\author{S.\ N.\ Coppersmith}
\email{snc@physics.wisc.edu}
\affiliation{%
Department of Physics, University of Wisconsin--Madison,
Madison, Wisconsin 53706, USA
}%
\affiliation{%
School of Physics, The University of New South Wales,
Sydney, New South Wales 2052,  Australia}
\author{Calvin W.\ Johnson}
\email{cjohnson@sdsu.edu}
\affiliation{%
 Department of Physics, San Diego State University, San Diego,
 California 92182-1233, USA
}

\date{\today}

\begin{abstract}
{We investigate the importance of going beyond the mean-field approximation in the dynamics of collective neutrino oscillations. To expand our understanding of the coherent neutrino oscillation problem, we apply concepts from {many-body physics} and quantum information theory. Specifically, we use measures of nontrivial correlations (otherwise known as ``entanglement'') between the constituent neutrinos of the many-body system, such as the entanglement entropy and the Bloch vector of the reduced density matrix.  The relevance of going beyond the mean field is demonstrated by comparisons between the evolution of the neutrino state in the many-body picture vs the mean-field limit, for different initial conditions. }
\end{abstract}

\maketitle

\section{Introduction} \label{section:intro}

Collective neutrino oscillations, which result from coherent forward scattering of neutrinos off other free-streaming neutrinos, is a topic of long-standing interest, particularly in connection with neutrinos coming from core-collapse supernovae and merging binary neutron stars, as well as neutrinos in the early universe~\cite{Fuller87, Notzold:1988fv, Pantaleone:1992eq, Sigl:1993fr, Raffelt:1993kx, Samuel:1993sf,Kostelecky:1993ys, Kostelecky:1995rz, Dolgov:2002ab, Ho:2005vj, McKellar:1994uq, Enqvist:1991yq, Pastor:2002zl, Abazajian:2002lr, Hannestad:2006qd, Johns:2016enc,Duan:2009cd, Duan:2010fr, Chakraborty:2016a, Balantekin:2018mpq, Duan06a, Duan06b, Duan06c, Duan07a, Duan07b, Duan07c, Duan:2008qy, Raffelt07, Raffelt:2007, Fogli07, Dasgupta:2007ws, Dasgupta:2008qy, Dasgupta09, Friedland:2010yq, Dasgupta:2010cd, Galais:2011gh, Mirizzi:2013rla, Mirizzi:2013wda, Pehlivan:2016lxx, Tian:2017xbr, Das:2017iuj, Birol:2018qhx, Cirigliano:2018rst, Malkus:2012, Malkus:2014, Wu:2016, Stapleford:2016jgz, Malkus:2016, Frensel:2016fge, Vaananen:2016, Zhu:2016mwa, Johns:2016enc, Shalgar:2017pzd, Vlasenko:2018irq,Sawyer:2005yg, Chakraborty:2016lct, Capozzi:2016oyk, Wu:2017qpc, Sen:2017ogt, Dighe:2017sur, Abbar:2017pkh, Wu:2017drk, Dasgupta:2017oko, Dasgupta:2018ulw, Capozzi:2018clo, Abbar:2018beu, Shalgar:2019kzy, Capozzi:2019lso} (see, e.g., reviews in Refs.~\cite{Duan:2010fr,Duan:2009cd,Chakraborty:2016a}). {Terrestrial experiments such as DUNE, with the capability to detect supernova neutrinos, could potentially be used to discern the signatures of collective oscillations in the observed energy spectra and flavor content of the neutrinos and thereby yield new insight into the nature of neutrinos and the physics of core-collapse supernovae.} 

It is not {practical} 
to determine  the exact evolution of $10^{58}$ or so neutrinos and antineutrinos emitted during the core collapse of a typical supernova. 
 For example, for a two-flavor system of $N$ neutrinos (with no antineutrinos), the dynamical variable, i.e., the $N$-particle wave function, occupies a $2^N$-dimensional Hilbert space that is the direct product of the individual one-body Hilbert spaces. 
{Therefore {it is common} to adopt simplifying approaches such as the mean-field approximation.}
This approach, however, is rather drastic;
{in the mean-field approximation, all quantum correlations between particles are ignored, ensuring that the dynamical variable is always separable into $N$ single-particle wave functions, thereby effectively reducing the size of the Hilbert space to just $2N$}. It is therefore crucial to test the limits of validity of the mean-field approximation in different scenarios. 

The size of the Hilbert space in an interacting many-body system scales exponentially with the number of particles; it is therefore computationally expensive to analyze differences between many-body and mean-field approaches in realistic supernova models with large numbers of neutrinos. In this work, we therefore adopt a toy model consisting of a few neutrinos, each having discrete well-defined momenta. This discretization necessitates considering the neutrinos as plane waves, and as a result the neutrino-neutrino interactions in our model are not localized in space.

Some previous treatments have used this approach, albeit with additional simplifications such as ignoring vacuum oscillations. For instance, it has been argued that, when the neutrinos start from an initial configuration that is  asymmetrically distributed in momentum space, collisions can cause them to become uniformly distributed faster than one would expect from single-particle cross sections, due to many-body entanglement~\cite{Bell:2003mg}. On the other hand, it was shown that, for certain geometries (i.e., orthogonal neutrino beams) in the absence of vacuum oscillations, the contribution from entangled states becomes insignificant if successive neutrino-neutrino scatterings are treated as independent events~\cite{Friedland:2003dv,Friedland:2003eh}. More recently, there have been attempts to characterize the neutrino many-body Hamiltonian in terms of its eigenvalues and eigenstates~\cite{Balantekin:2007kx,Pehlivan:2011hp,Birol:2018qhx,Patwardhan:2019zta}, and this insight has been applied to study the evolution of such a system for a particular initial condition~\cite{Birol:2018qhx}.

As an aside, previous authors have also explored entanglement of neutrinos with charged leptons in particle decays~\cite{Kayser:2010bj, Boyanovsky:2011xq, Floerchinger:2018mla} and in electron capture~\cite{Pavlichenkov:2011vk}, flavor entanglement during both free propagation and interaction with the background environment~\cite{Blasone:2007vw, Akhmedov:2010ua, Wu:2010yr, Blasone:2015lya, Ciuffoli:2019qdx}, {as well as entanglement between a two-level system (which could model two neutrino flavors) and a thermal bath (which could represent the environment)~\cite{bell2002entanglement}}. However, in these cases, neutrino propagation is treated as a one-body problem from the perspective of the neutrino, in contrast with collective neutrino oscillations which arise from neutrino-neutrino interactions and represent a many-body problem. 

{In this work we focus on quantifying the entanglement that can develop in a many-body neutrino system as described by our toy model.
{For this purpose, we calculate the time evolution of the entanglement entropy of the different neutrinos in systems with varying sizes and initial configurations.} Since entanglement is absent in mean-field treatments, an entanglement measure can serve as a quantifier for the extent to which many-body systems can deviate from the mean-field approximation. Indeed, we observe that the entropy of entanglement {appears to be correlated with the magnitude of the differences in the flavor evolution between {many-body} and mean-field results, with significant deviations from the mean field observed in some cases}.

In our analysis we consider adiabatic evolution of the many-neutrino system, with at most one neutrino in each energy bin, in the single-angle approximation. The equations defining the model that we study are presented in Sec.~\ref{sec:model}.
Adiabatic eigenvalues and eigenstates of this many-neutrino Hamiltonian {are} obtained~\cite{Pehlivan:2011hp, Birol:2018qhx, Patwardhan:2019zta} using the Richardson-Gaudin algebraic approach~\cite{Richardson:1966zza, Gaudin:1976sv, Balantekin:2018mpq}. This solution is outlined briefly in Sec.~\ref{section:Evolution}. 
In Sec.~\ref{section:GenEntangle} we define the entanglement measures that we have used in our analysis. Section~\ref{section:OverResults} presents an overview of the main results, in particular the relationship between entanglement entropy and the deviation of the many-body results from their mean-field counterparts, for small-$N$ systems up to $N=9$. Section~\ref{section:N=2} {presents} the solution for a two-neutrino system, which can be obtained analytically. A short summary of the mean-field evolution equations for two neutrinos is also given. Section~\ref{section:N>2} contains {detailed} numerical solutions for the cases where the number of neutrinos is greater than two. 
Finally, Sec.~\ref{section:Conclusion} contains a brief discussion of our conclusions. In Appendix~\ref{Appendix_Eigen}, we summarize our procedure for obtaining eigenvalues and eigenvectors of the single-angle neutrino Hamiltonian, and in Appendix~\ref{Appendix_Ev} we explore the validity of using the adiabatic approximation for evolving the many-body system. 

\section{The Neutrino Hamiltonian} \label{sec:model}

We consider the evolution of an ensemble of $N$ free-streaming neutrinos undergoing vacuum and collective oscillations in two flavors: $e$ and $x$. We neglect inelastic collisions between these neutrinos, an approximation that may not be appropriate under all circumstances~\cite{Cherry:2012lr} but has the advantage of yielding a {more} tractable model. This system {is} described by the many-body Hamiltonian \cite{Pehlivan:2011hp}
\begin{equation}
	H = \sum_{\mathbf{p}}\omega_\mathbf{p}\vec{B}\cdot\vec{J}_\mathbf{p} + \sum_{\mathbf{p},\mathbf{q}}\mu_{\mathbf{p}\mathbf{q}}\vec{J}_\mathbf{p}\cdot\vec{J}_\mathbf{q}~,
	\label{multiH}
\end{equation}
where the vacuum oscillation frequencies are $\omega_\mathbf{p}=\delta m^2/2|\mathbf{p}|$ and {$\mathbf{p}$ are the neutrino momenta}. The strength of collective interactions for each pair of neutrinos is 
\begin{equation}
	\mu_{\mathbf{pq}} = \frac{\sqrt{2}G_F}{V}(1-\cos\theta_{\mathbf{pq}}),
	\label{multiMu}
\end{equation}
obtained from the leading-order effective four-point Fermi weak interaction diagrams of neutrinos in two flavors; $G_F$ is the Fermi coupling constant, $V$ is the volume in a box quantization, and $\theta_{\mathbf{pq}}$ is the angle between the momenta of interacting neutrinos. 
The weak isospin vectors $\vec{J}_\mathbf{p}$ for each neutrino are defined in terms of creation and annihilation operators of individual neutrinos in the mass basis: $J_\mathbf{p}^z=\frac{1}{2}(a_{1,\mathbf{p}}^\dagger a_{1,\mathbf{p}}^{\phantom{\dagger}}-a_{2,\mathbf{p}}^\dagger a_{2,\mathbf{p}}^{\phantom{\dagger}})$, $J_\mathbf{p}^+=a_{1,\mathbf{p}}^\dagger a_{2,\mathbf{p}}^{\phantom{\dagger}}$, and $J_\mathbf{p}^-=a_{2,\mathbf{p}}^\dagger a_{1,\mathbf{p}}^{\phantom{\dagger}}$. Here {we define} isospins up and down {as} correspond{ing} to {the neutrino vacuum mass eigenstates $\nu_1$ and $\nu_2$}. {The} vector $\vec{B}$, {which characterizes the oscillations of the individual neutrinos and plays a role similar to that of an external magnetic field if the $J$'s are interpreted as spins}, takes the value $\vec{B} = (0,0,-1)$ in the mass basis. 
Note that each of the isospin operators and the vector $\vec{B}$ may also be expressed in the flavor basis, using a two-flavor mixing angle, $\theta$, unrelated to the $\theta_{\mathbf{pq}}$ in Eq.~\eqref{multiMu}.


{We study the tractable, but nevertheless interesting problem
that is obtained by taking a} 
geometric average over the angles $\theta_{\mathbf{pq}}$, {called the single-angle approximation}, {which results in}
a single collective interaction strength as a function of position or time. For example, 
in a spherical neutrino bulb model~{\cite{Duan06a,Duan:2010fr}},
\begin{equation}
	\mu(r) = \frac{G_F}{\sqrt{2}V}\bigg[1-\bigg(1-\frac{R_\nu^2}{r^2}\bigg)^{1/2}\bigg]^2,
	\label{singleMu}
\end{equation}
 where $r$ is the distance from the center of a neutrino sphere of radius $R_\nu$. Here we adopt this form of $\mu(r)$ for the rest of our study. At this point, we can replace the $3$-vector momenta $\mathbf{p}$ with simple indices $p=1,\ldots,M$ to label the {discrete energy bins in our model}. {For brevity, we henceforth also refer to $\mu(r)$ as simply $\mu$}, except when the $r$ dependence is relevant for calculations. Thus, our Hamiltonian reduces to the simpler form
\begin{equation}
	H = -\sum_{p=1}^M\omega_pJ_p^z + \mu\vec{J}\cdot\vec{J},
	\label{singleH}
\end{equation}
where $\vec{J}=\sum_{p=1}^M\vec{J}_p$. This simpler Hamiltonian has several conserved quantities that may be helpful to consider, such as the total $z$-component isospin $J^z=\sum_{p=1}^MJ_p^z$ and a set of $M$ conserved charges \cite{Pehlivan:2011hp}:
\begin{equation}
	h_p=-J_p^z+2\mu\sum_{\substack{q=1\\q\neq p}}^M\frac{\vec{J}_p\cdot\vec{J}_q}{\omega_p - \omega_q}.
	\label{charges}
\end{equation}
Interestingly, the Hamiltonians in Eqs.~\eqref{singleH} and~\eqref{charges} are also of interest in condensed matter physics, e.g., Refs.~\cite{Owusu_2008, Faribault:2011rv, Claeys:2017sp}. 

For simplicity, {we choose a system where the $N$ neutrinos are distributed evenly across $M$ oscillation frequencies with $N=M$; therefore} $j_p=1/2$ for $p=1,\ldots,N$. For this case, methods were {developed} to obtain all $2^N$ solutions, with eigenvalues and eigenstates of the Hamiltonian and these conserved quantities determined numerically for $N\geq2$ (for details, refer to Appendix~\ref{Appendix_Eigen} or Refs.~\cite{Claeys:2018zwo,Cervia:2019nzy,Patwardhan:2019zta}). These results are recorded as functions of $\mu/\omega_0$, with the oscillation frequencies fixed, $\omega_p=p\omega_0$; our value of $\omega_0$ is given in {Sec.~\ref{section:N>2}, Table~\ref{parameters}}.

\begin{table}
\caption{\label{parameters}Summary of parameters used in our calculations.\\}
\begin{ruledtabular}
\begin{tabular}{lcr}
 	&	\multicolumn{ 2}{c}{{Value}} \\ \cline{2-3}
{Parameter} & \multicolumn{1}{c}{Number} & \multicolumn{1}{c}{Unit}	\\ 
\colrule
$R_\nu$		&	$60$	&	km	\\
$\omega_0$  &	$1.055\times10^{-16}$	&	MeV	\\
$\mu(R_\nu)$    & $3.62 \times10^{4}$   &  $ \omega_0$       \\
$\sin^2(2\theta)$	&	$0.10$		&				\\
\end{tabular}
\end{ruledtabular}
\end{table}



\section{Adiabatic eigenstates and eigenvalues of the Hamiltonian} \label{section:Evolution}



The Hamiltonian in Eq.~\eqref{singleH} has eigenstates $\ket{q}$ where $q=0,1,\ldots,2^N-1$, which we may determine as superpositions of Kronecker products of mass basis states:
\begin{equation}
	\ket{q} = \sum_{i_1,\ldots,i_N=1}^2 V_{i_1,\ldots,i_N}(\mu) \ket{\nu_{i_1},\ldots,\nu_{i_N}}
	\label{genericEstate}
\end{equation}
where the $2^N$ coefficients $V_{i_1,\ldots,i_N}(\mu)$ for each state are functions of $\omega_p$ and $\mu$.\footnote{Here, we make the dependence of these coefficients on $\mu$ explicit, since we change its value as time progresses. In contrast, we fix our values of $\omega_p$ at the beginning of our problem, and they do not change as time progresses.} These coefficients have simple forms at $\mu=0$: $V_{i_1,\ldots,i_N}(0) = \prod_{j=1}^N\delta_{i_j q_j}$, where 
$q=\sum_{j=1}^N(q_j-1)\,2^{N-j}$ is the binary representation of $q$ (i.e., $q_j - 1$ is the $j$th digit from the left).
That is, the energy eigenstates completely reduce to products of mass eigenstates of the individual neutrinos in the $\mu\to0$ limit of the Hamiltonian in Eq.~\eqref{singleH}, as one would expect. For generic $\mu>0$, we begin with these solutions and numerically solve a system of $N$ coupled quadratic equations that constrain these coefficients. The relationship between these equations and coefficients [as well as eigenvalues of the charges in Eq.~\eqref{charges}] is described in Appendix \ref{Appendix_Eigen}. From these equations one can find that all of these coefficients must be real valued. Therefore, the $2^N$ sets of $2^N$ coefficients compose a $2^N\times2^N$ real matrix $V$ mapping energy states to mass states for a fixed $\mu$.

Suppose $\ket{\Psi}$ is the wave function of a many-body system with a Hamiltonian $H$, both in the mass basis. Then the time evolution of the system is given by
\begin{equation}
i \diff{\ket\Psi}{t} = H\ket\Psi.
\end{equation}

The Hamiltonian may be diagonalized as $H = V \Sigma V^\dagger$ by a unitary transformation $V$, and one may define $\ket\Psi = V \ket\Phi$ to rewrite the equation as
\begin{equation}
i \diff{\ket\Phi}{t} + i V^\dagger \diff{V}{t} \ket\Phi = \Sigma \ket\Phi.
\end{equation}

Assuming adiabaticity,\footnote{For a few initial conditions, we compared our adiabatic many-body results with the corresponding solutions obtained without assuming adiabaticity, and confirmed that they agree up to a reasonable level of precision (comparison not shown here).} i.e., neglecting the second term on the left-hand side, one obtains an evolution equation for $\ket\Phi$, which may be integrated to obtain

\begin{equation}
\ket{\Phi(t)} = e^{-i\int^t \Sigma dt'} \ket{\Phi_0}.
\label{basicTimeEvolve}
\end{equation}

{Since} $\Sigma$ is a diagonal matrix, it is easy to exponentiate. 
In Appendix~\ref{Appendix_Ev} we argue the time evolution operator used here does not need to be a Dyson series, as the commutator of Hamiltonians at different time values is {$[H(t),H(t')]\sim\mathcal{O}(t-t')$} and therefore is numerically insignificant for small time steps. The equation for $\ket\Psi$ then becomes
\begin{equation}
\ket{\Psi(t)} = V e^{-i\int^t \Sigma dt'} V_0^\dagger \, \ket{\Psi_0} {,}
\end{equation}
where $V$ and $V_0$ are the unitary {transformations} {that diagonalize the Hamiltonian} at {the values} $\mu=\mu(r)$ and the initial value $\mu_0\equiv\mu(r_0)$, respectively, with $r_0\geq R_{\nu}$, where $R_\nu$ is the neutrino sphere radius as introduced in Eq.~\eqref{singleMu}.
Starting from an initial state $\ket{\Psi(\mu_0)}$ of our many-body neutrino system,
the adiabatic evolution of the system is then given by
\begin{align}
	\ket{\Psi(\mu)} =& V \bigg[\sum_{q=0}^{2^N-1}\exp\bigg({-i\int_{\mu_0}^\mu \Sigma_{qq}(\mu')\frac{dr}{d\mu'}\mathrm{d}\mu'}\bigg) 
	\ket{q}\!\bra{q}\bigg]
 	\nonumber \\
	& \times V_0^T\ket{\psi(\mu_0)}
	\label{evolvedState}
\end{align}
where $\Sigma_{qq}\equiv \braket{q|\Sigma|q}$. 
Note that here we have used the position $r$ as a proxy for time $t$ in the evolution equation, as is justified if we assume a steady-state configuration wherein the interaction strength $\mu$ depends explicitly only on position and not time.

It is worth pointing out, as observed in Refs.~\cite{Birol:2018qhx, Patwardhan:2019zta}, that energy eigenvalues of the Hamiltonian in Eq.~\eqref{singleH} exhibit numerous energy level crossings. However, in our previous study~\cite{Patwardhan:2019zta}, it was determined that the conserved charge operators in Eq.~\eqref{charges} cannot all be simultaneously degenerate for any given value of $\mu$. Therefore, as is argued in Appendix~\ref{Appendix_Ev}, we propose that one can adiabatically evolve this state without encountering nontrivial level crossings to be dealt with, as these charges break any degeneracies that may arise in our Hamiltonian.

\section{Quantifying entanglement}
\label{section:GenEntangle}

{In this section we review the entanglement entropy and
the Bloch vector, which are well-established useful concepts in the context of quantum information theory~\cite{ Everett:1956dja, Nielsen:2011:QCQ:1972505}. }
For an $N$-particle wave function $\ket{\Psi}$ defined on a Hilbert space $\mathcal{H} = \mathcal H_1 \otimes \cdots \otimes \mathcal H_N$, one can define the density matrix $\rho = \ket{\Psi}\!\bra{\Psi}$. Partitioning the Hilbert space as $\mathcal{H} = \mathcal H_A \otimes \mathcal H_B$, one can obtain the reduced density matrix for partition $A$ by taking the partial trace over partition $B$, 
\begin{align}
\rho_A	&= \mathrm{Tr}_{B}[\rho].
\label{GeneralReducedDensity}
\end{align}

A suitable measure of entanglement for nonmixed states is the bipartite entanglement entropy. For partitions $A$ and $B$, the entanglement entropy is defined as
\begin{equation}
S_{(A|B)} = -\text{Tr}[\rho_A \log \rho_A] = -\text{Tr}[\rho_B \log \rho_B] ,
\label{EntEntropy}
\end{equation}

In this work, we choose one of the neutrinos to be in partition $A$, and the remaining neutrinos to be in partition $B$, so that the reduced density matrix $\rho_A$ is simply a $2 \times 2$ matrix. For example, if we choose the neutrino at the highest $\omega$ to be in partition $A$, then $\rho_A$ is given by
\begin{align}
\rho_A = \sum_{i_{1},\ldots,i_{N-1}=1,2} \braket{\nu_{i_{1}}\ldots\nu_{i_{N-1}}|\rho|\nu_{i_{1}}\ldots\nu_{i_{N-1}}}
\label{densityRed}
\end{align}

One may define the \lq\lq polarization vector\rq\rq\ for each particle as $\vec P_\omega = 2\braket{\Psi|\vec J_\omega|\Psi}$. If $\mathcal{H}_A$ contains a single neutrino, its polarization vector $\vec P_A$ is related to its reduced density matrix as 
\begin{equation}
    \rho_A = \frac12 (\mathbf I + \vec{\sigma}\cdot\vec{P}_A),
    \label{polarVec}
\end{equation}
where $\mathbf{I}$ is the $2\times2$ identity matrix and $\vec\sigma$ are the Pauli spin matrices. For the sake of brevity, we drop the subscripts ``$A$'' and ``$(A|B)$'' from the polarization vector and entropy, respectively, in the remainder of the text. In this scenario, the polarization vector is simply the standard $SU(2)$ Bloch vector. It is straightforward to check that the eigenvalues of $\rho_A$ are $(1\pm P)/2$, where $P \equiv|\vec{P}|$. One can then relate the magnitude $P$ of the polarization vector to the entropy of entanglement via the relation
\begin{equation}
    S = - \frac{1-P}{2} \log \left( \frac{1-P}{2} \right) -  \frac{1+P}{2}  \log \left(\frac{1+P}{2} \right).
    \label{concEntropy}
\end{equation}
In this sense, $P$ may be used as a measure of entanglement; $P = 1$ indicates a pure state, whereas $P < 1$ implies that the reduced density matrix $\rho_A$ is mixed, signaling entanglement between the two partitions. Additionally, one can observe that $P$ is related to the concurrence of the state across the partitions $A$ and $B$, via $C=\sqrt{1-P^2}$~\cite{wootters1998entanglement,hill1997entanglement}.




\section{Overview of the results} \label{section:OverResults}

Here we examine the differences between the predictions of mean-field theory and many-body results obtained for small $N$ and also show that the magnitude of these differences is strongly correlated with measures of quantum entanglement in the many-body solutions. To explore differences between the mean-field approximation and the adiabatic many-body solutions we calculated several quantities in both approaches. Details of these calculations are given in {subsequent} 
sections. In this section we provide an overview of our results. The key results that we obtain are the following:
\begin{enumerate}
\item  The magnitude of the differences between the results obtained using the mean-field approximation and those obtained using the many-body approach are correlated with measures of entanglement of the asymptotic configuration reached in the limit of long times, and
\item Deviations between the many-body solutions and mean-field theory for asymptotic values of observables
depend strongly on the initial condition and often are substantial. 
\end{enumerate}

To quantify physically relevant differences between the mean-field and many-body solutions, in Fig.~\ref{fig:pzwn} we compare asymptotic values (i.e., at large radius $r$ where $\mu \ll \omega_0$) of the $z$ component of the polarization vector, $P_z$, for the neutrino with the highest frequency $\omega_N = N\omega_0$. The asymptotic values are calculated in both the mean-field and the many-body approaches, for different values of $N$, the number of neutrinos. In this figure the initial state is a state with $N$ electron neutrinos. For this initial configuration{, the} deviation {between the many-body and the mean-field results} increases with the number of neutrinos. {Furthermore, in Fig.~\ref{GeneralEntropyR} we present the evolution of the entropy of entanglement $S$ of the highest frequency neutrino, with the rest of the ensemble [calculated using Eqs.~\eqref{EntEntropy} or \eqref{concEntropy}]. Note that the entanglement entropy is always zero in the mean-field approximation. The growth in asymptotic values of $S$ with $N$ demonstrates that the entanglement entropy and the {magnitude of} {the discrepancy of $P_z$} {between the exact and mean-field solutions} are correlated.}
\begin{figure*}
\begin{center}
    \subfloat[\label{fig:pzwn}]{
	    \includegraphics[width=0.48\textwidth]{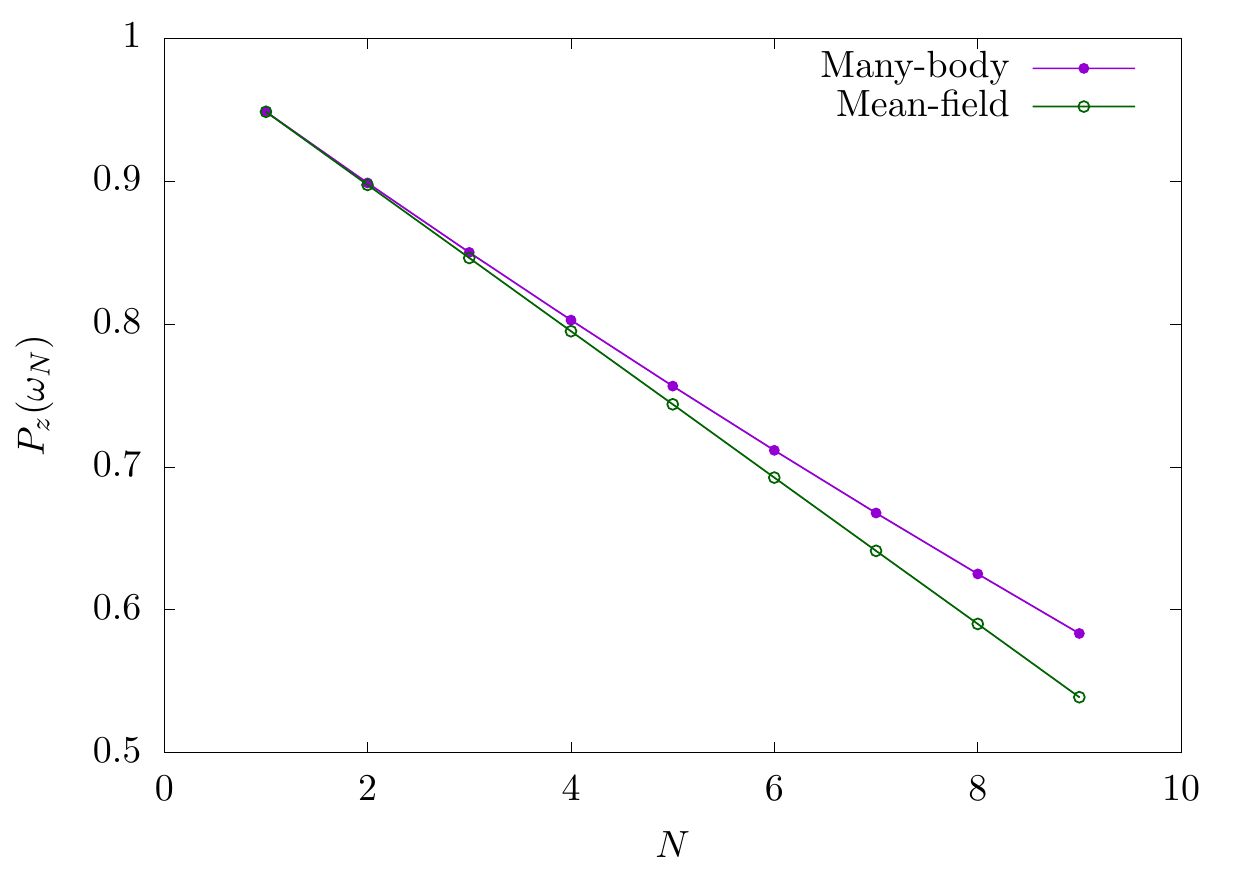}}
	~
	\subfloat[\label{GeneralEntropyR}]{
	    \includegraphics[width=0.48\textwidth]{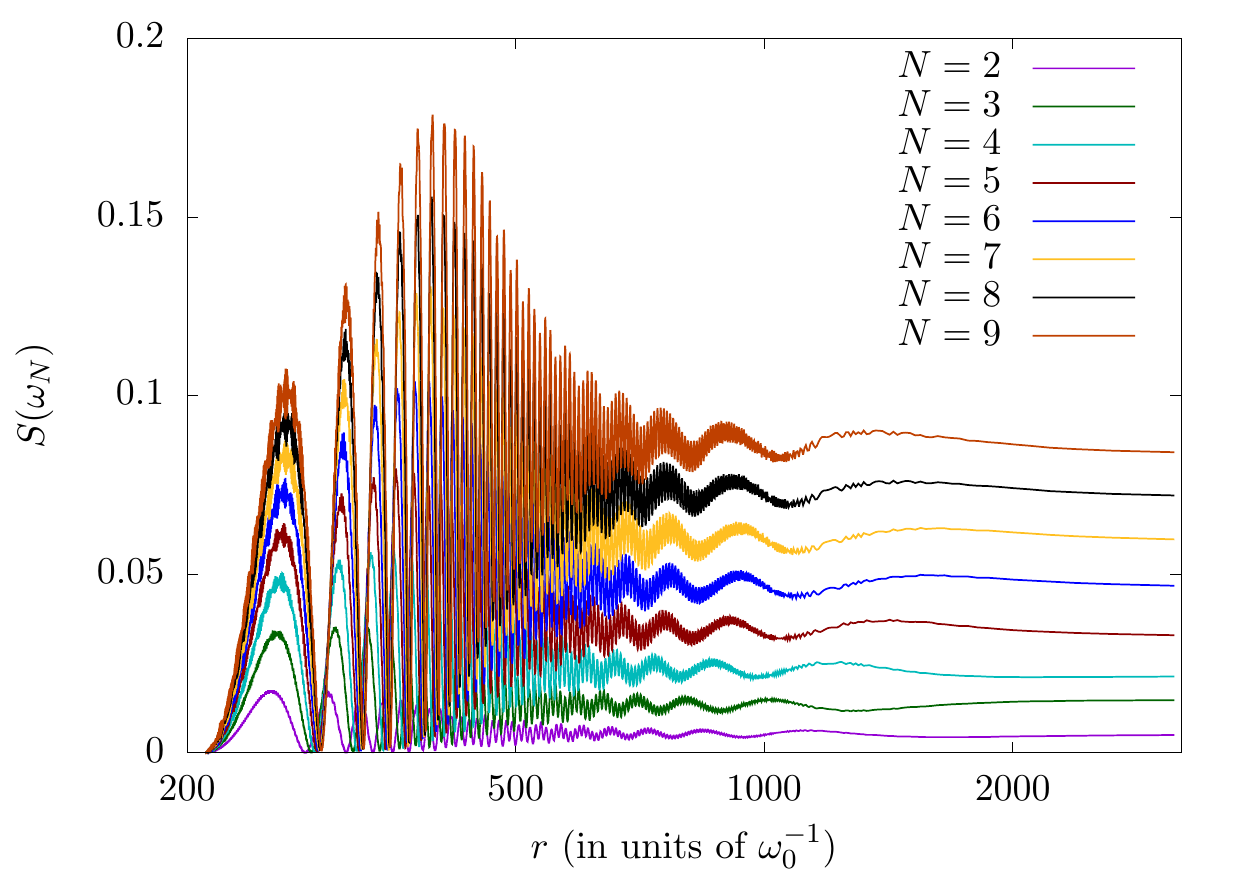}}
    
\end{center}
    \caption{Left: Comparison between mean-field and many-body calculations of the $z$ component of the polarization vector, $P_{z}(\omega_N)$, for the neutrino with highest frequency $\omega_N$. Shown here are the asymptotic (i.e., at $r\gg R_\nu$) values of $P_{z}(\omega_N)$, for systems with various numbers of neutrinos, $N$, each starting from an initial configuration $\ket{\nu_e, \ldots, \nu_e}$ at $\mu_0 = 5\omega_0$ [or, equivalently, at the radius $r_0\approx 6.6 R_\nu$, as per Table~\ref{parameters} and Eq.~\eqref{singleMu}]. Notably, for this configuration, the discrepancy between the mean-field and many-body results grows with increasing $N$, at least for the small values of $N$ examined here. Right: The entanglement entropy $S(\omega_N)$ of the highest frequency neutrino with the rest of the ensemble [calculated using Eqs.~\eqref{evolvedState}--\eqref{densityRed}], as a function of radius $r$, for the same set of systems as in the left panel. 
    In mean-field theory this entropy is zero. 
    Here the asymptotic values of $S(\omega_N)$ also 
    grow with $N$, demonstrating a possible correlation with the discrepancy between many-body and mean-field results.
    Note that $\mu$ decreases with $r$---here we use the relationship from Eq.~\eqref{singleMu}, which is borrowed from the single-angle bulb model~\cite{Duan:2010fr}.
    } 
    \label{fig:Pz&S_alle}
\end{figure*}

%
We next present results demonstrating that the many-body quantum correlations can vary greatly between different initial configurations. To illustrate this in  Fig.~\ref{fig:PvS} we display asymptotic ($r \gg R_{\nu}$) values of the difference 
\begin{equation}
\Delta P_z(\omega_p) \equiv |P_z^{\mathrm{MF}}(\omega_p)-P_z^{\mathrm{MB}}(\omega_p)|  
\end{equation}
$p= 1, \ldots, N$, where MF and MB denote mean-field and many-body results, respectively, for the $z$ component of the polarization vector of the neutrino with frequency $\omega_p = p \omega_0$, versus the entanglement entropy for this neutrino with the rest of the ensemble in the many-body calculation. The left panel is for $N=4$ neutrinos for all 16 ($2^4$) possible initial state configurations with neutrinos of definite flavor.\footnote{A global inversion of both flavor and the ordering of $\omega$ values of the evolved state results in another solution for a different initial configuration, with $\vec{P}(\omega_p)\mapsto-\vec{P}(\omega_{N-p})$. Therefore, each point appearing in Fig.~\ref{fig:PvS4}, in fact, represents the same data for two different initial configurations of overall opposite flavor and reversed ordering; for example, initial states $\ket{\nu_e\nu_x\nu_x\nu_x}$ and $\ket{\nu_e\nu_e\nu_e\nu_x}$ correspond to the same four data points $\{S(\omega_p),\Delta P_z(\omega_p)\}$ for $p=1,\ldots,4$.}
The right panel is for an $N=8$ neutrino system, for a subset of configurations where four of the neutrinos, placed at either $\omega_1,\ldots,\omega_4$ or $\omega_5,\ldots,\omega_8$ are iterated through the same 16 initial configurations as in the left panel, whereas the remaining ones are all taken to be in the $\nu_e$ flavor initially. This figure illustrates that, whenever a large deviation from mean-field theory is exhibited, the entanglement entropy tends to be large. However, the converse is not necessarily true---in some cases, even when the entanglement entropy is large, the deviations from mean-field theory can nevertheless be small. 
{The lines drawn in Fig.~\ref{fig:PvS} are of the quantity
\begin{equation}
\Delta P_z(\omega_p) = 1 - P(S(\omega_p)),    
\end{equation} 
where $P(S)$ is the inverse of the function in Eq.~\eqref{concEntropy}.}
We observe {that} a substantial number of $\Delta P_z(\omega_p)$ {values} cluster around this line, particularly for $N=8$. This alignment could occur, for instance, when $P_x$ and $P_y$ components of these vectors are vanishingly small.

\begin{figure*}[htb]
\begin{center}
    \subfloat[$N=4$\label{fig:PvS4}]{
	    \includegraphics[width=0.49\textwidth]{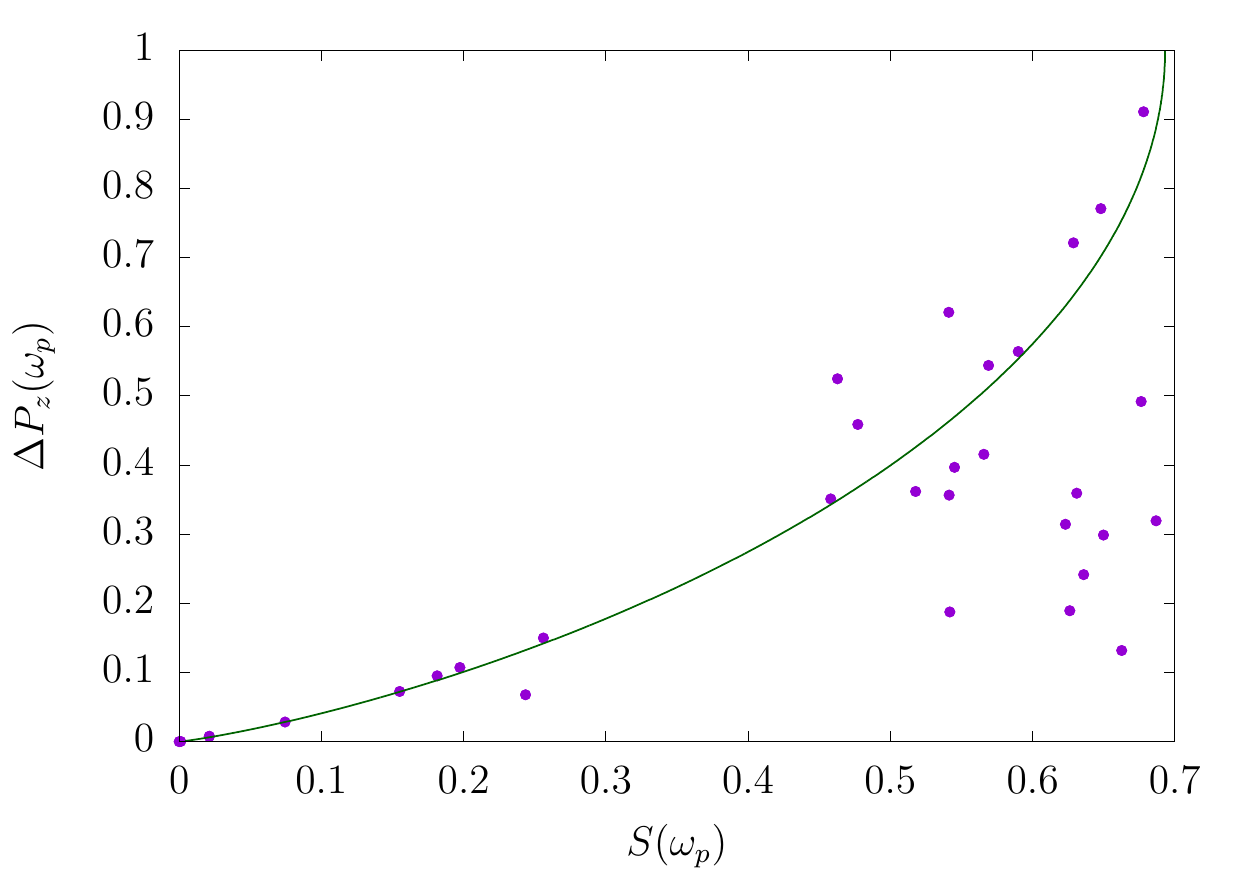}}
	~
	\subfloat[$N=8$\label{fig:PvS8}]{
	    \includegraphics[width=0.49\textwidth]{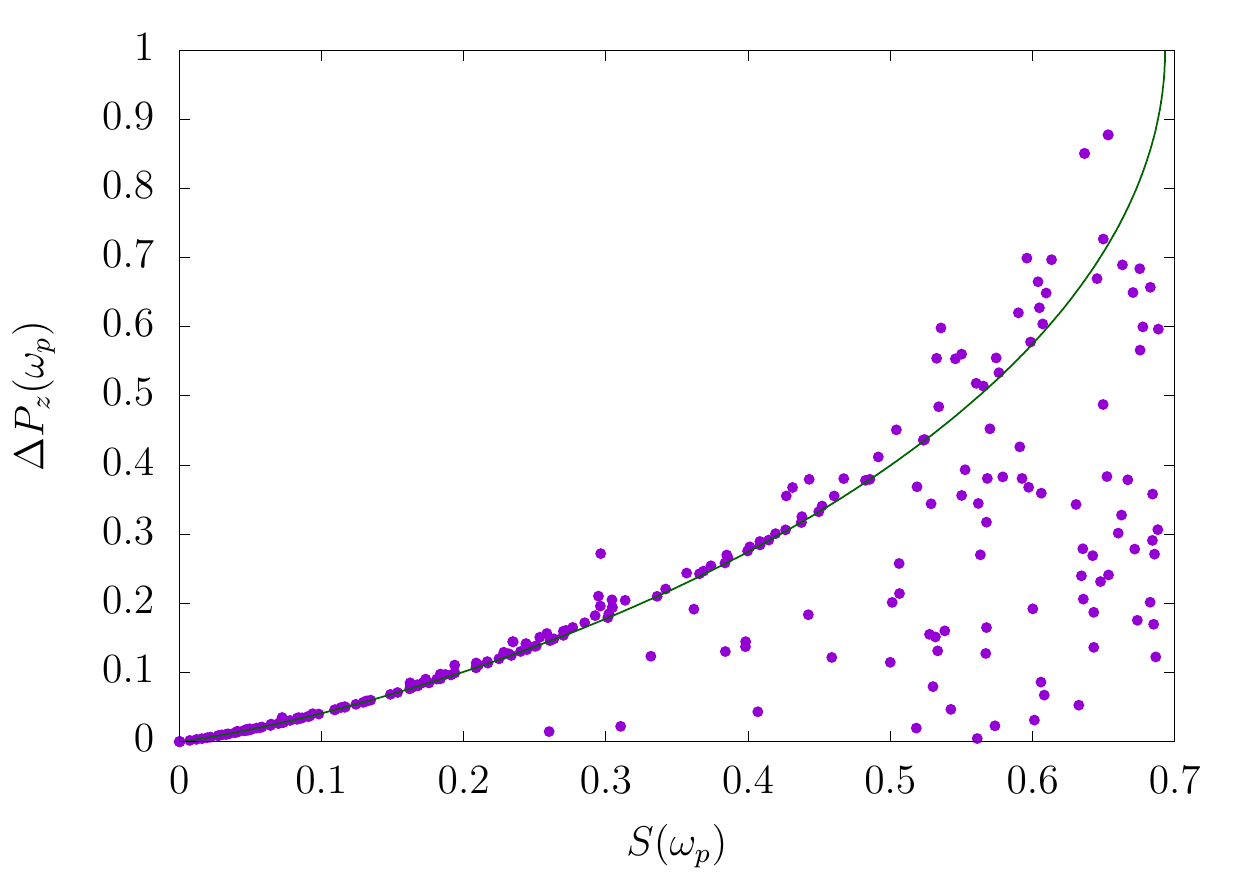}}
\end{center}
	\caption{Deviation between the asymptotic (i.e., at $r \gg R_\nu$) mean-field and many-body values of the $z$ component of the polarization vectors, $\Delta P_{z}(\omega_p) = |P_z^\mathrm{MF}(\omega_p) - P_z^\mathrm{MB}(\omega_p)|, \, p=1,\ldots,N$, versus the entanglement entropies $S(\omega_p)$, for various initial configurations. Each dot represents a particular neutrino within a system that starts from one of the chosen initial configurations. 
	Left: Dots showing asymptotic values of $\Delta P_z(\omega_p)$, for neutrinos starting from all 16 possible initial configurations of a system with $N=4$ (where each neutrino can start with a definite flavor $\nu_e$ or $\nu_x$). 
	Right: Dots showing asymptotic $\Delta P_z(\omega_p)$ values of neutrinos starting from a subset of possible initial configurations for $N=8$. The configurations included here consist of four of the neutrinos, with frequencies of either $\omega_1,\ldots,\omega_4$ or $\omega_5,\ldots,\omega_8$, being iterated through the same 16 initial configurations as in the left panel, and the remaining four neutrinos are all taken to be $\nu_e$ initially. 
	The correlation between $\Delta P_z$ and $S$ appears to strengthen with increasing $N$. Also shown in both figures is the curve $1 - P(S)$, where $P(S)$ is the inverse of the function in Eq.~\eqref{concEntropy}. The clustering  of the dots 
	around this line is discussed in the text.}
	\label{fig:PvS}
\end{figure*}

\section{Analytic results for a two-neutrino system} \label{section:N=2}

{Before we {elaborate} further on the results overviewed in Sec.~\ref{section:OverResults}, it is instructive to illustrate the example of the two-neutrino system. This system is simple enough to be examined analytically, but it nevertheless brings to light some key features that are generically present in many-body systems. In Sec.~\ref{sec:adiabaticN2}, we describe the adiabatic evolution of the $N=2$ neutrino system starting from the different possible initial conditions, and in Sec.~\ref{sec:N=2mft}, for comparison, we present the mean-field evolution equations for the same system.}

\subsection{Many-body results in the adiabatic limit} \label{sec:adiabaticN2}

For two neutrinos, the Hamiltonian in the product mass basis can be written as
\begin{equation}
H = \begin{pmatrix}
-\Omega + 2\mu & 0 & 0 & 0 \\
0 & -\eta + \mu & \mu & 0 \\
0 & \mu & \eta + \mu & 0 \\
0 & 0 & 0 & \Omega + 2\mu 
\end{pmatrix}
\end{equation}
where $\Omega\equiv(\omega_1+\omega_2)/2$ and $\eta\equiv(\omega_1-\omega_2)/2$. This can be diagonalized into 
\begin{equation}
\Sigma = \begin{pmatrix}
-\Omega + \mu & 0 & 0 & 0 \\
0 & \sqrt{\eta^2 + \mu^2} & 0 & 0 \\
0 & 0 & -\sqrt{\eta^2 + \mu^2} & 0 \\
0 & 0 & 0 & \Omega + \mu 
\end{pmatrix},
\end{equation}
where we have also subtracted a term {\color{black} $\mu\mathbb{I}_{4\times4}$, which contributes only a global phase to the evolved state and therefore is not present in calculations with the resulting density matrix of the state.} Here, the unitary transformation matrix can be parametrized as 
\begin{equation}
V = \begin{pmatrix}
1 & 0 & 0 & 0 \\
0 & \cos{\xi} & -\sin{\xi} & 0 \\
0 & \sin{\xi} & \cos{\xi} & 0 \\
0 & 0 & 0 & 1 
\end{pmatrix},
\end{equation}
where one has 
\begin{equation}
\cos{2\xi} = \frac{-\eta}{\sqrt{\mu^2 + \eta^2}}    
\>\>\> \>
{\rm and} 
\>\>\>\sin{2\xi} = \frac{\mu}{\sqrt{\mu^2 + \eta^2}}.\label{xiConstraints}
\end{equation}
One can thus observe that only the second and the third components of the wave function mix. Writing the wave function as $\ket{\Psi(t)} = (\alpha(t), \beta(t), \gamma(t), \delta(t))^{T}$, one obtains the following evolution equations in the adiabatic limit:
\begin{align}
\alpha(t) = &e^{i\Omega t - iR} \alpha_0 \\
\beta(t)  = &\cos Q \, [\beta_0 \cos(\xi-\xi_0) - \gamma_0 \sin(\xi-\xi_0)] \nonumber \\
            &- i\sin Q \, [\beta_0 \cos(\xi+\xi_0) + \gamma_0 \sin(\xi+\xi_0)] \\
\gamma(t) = &\cos Q \, [\beta_0 \sin(\xi-\xi_0) + \gamma_0 \cos(\xi-\xi_0)] \nonumber \\
            &- i\sin Q \, [\beta_0 \sin(\xi+\xi_0) - \gamma_0 \cos(\xi+\xi_0)] \\
\delta(t) = &e^{-i\Omega t - iR} \delta_0 , 
\end{align}
where $\alpha_0,\ldots,\delta_0$ and $\xi_0$ are the respective initial values, and where $R = \int^t_{t_0} \mu\, dt'$ and $Q = \int^t_{t_0} \sqrt{\mu^2 + \eta^2} \, dt'$. The reduced density matrix for the {second} particle can be written as 
\begin{equation} \label{eq:rhored}
\rho_{2}^\text{red} = \begin{pmatrix}
|\alpha|^2 + {|\gamma|^2} & \alpha {\beta^*} + {\gamma} \delta^* \\
\alpha^* {\beta} + {\gamma^*} \delta & {|\beta|^2} + |\delta|^2
\end{pmatrix}.
\end{equation}
For {this subsystem}, one can use trigonometric identities to obtain an expression for $P_z = (\rho_2^\text{red})_{11} - (\rho_2^\text{red})_{22}$,
\begin{equation}
\begin{split}
P_z =& |\alpha|^2 {-} |\beta|^2 {+} |\gamma|^2 - |\delta|^2 \\
    =& \alpha_0^2 - \delta_0^2 {-}  (\beta_0^2 - \gamma_0^2)
    \nonumber \\
    &\times[\cos 2\xi \cos 2\xi_0 + \sin 2\xi \sin 2\xi_0 \cos 2Q] 
    \nonumber \\
    &{-} 2 \beta_0\gamma_0  [ \cos 2\xi \sin 2\xi_0 - \sin 2\xi \cos 2\xi_0 \cos 2Q ]{.}
	\label{Pz2general}
\end{split}
\end{equation}

One can, in principle, also obtain analytic expressions for $P_x = 2\,\mathfrak{Re}[(\rho_2^\text{red})_{12}]$ and $P_y = -2\,\mathfrak{Im}[(\rho_2^\text{red})_{12}]$. Of particular interest, however, is the magnitude $P$, since it is directly related to the entanglement entropy, as described previously {with Eq.~\eqref{concEntropy}}. First, using Eq.~\eqref{eq:rhored}, one can write down an expression for $P_\perp = |P_x - i P_y| = 2 |(\rho_2^\text{red})_{12}|$, given by
\begin{equation}
{P_\perp = 2|\alpha^*\beta+\gamma^*\delta|.}
\end{equation}
These may then be combined with $P_z$ to obtain a succinct expression for the magnitude $P$, given by
\begin{align}
    P^2 =& 1 - 4|\alpha\delta-\beta\gamma|^2
    \nonumber \\
    =& 1 - [-2\alpha_0\delta_0\sin2R+(\beta_0^2+\gamma_0^2)\sin2\xi\sin2Q\big]^2 
    \nonumber \\
    &- [2\alpha_0\delta_0\cos2R+(\beta_0^2-\gamma_0^2)
    \nonumber \\
    &\times (\sin2\xi_0\cos2\xi-\cos2\xi_0\sin2\xi\cos2Q)
    \nonumber \\
    &-2\beta_0\gamma_0(\cos2\xi_0\cos2\xi+\sin2\xi_0\sin2\xi\cos2Q)]^2 {~.}
    \label{Pmag}
\end{align}
Alternatively, this expression can {be} obtained by recognizing that the determinant of $\rho_B$ is the product of its eigenvalues, and is therefore given by $(1-P^2)/4$ (as per our discussion in Sec.~\ref{section:GenEntangle}).

One can now examine the evolution of this system starting from different initial conditions. For a system where the initial state is $| \nu_{e} \nu_{e} \rangle$ or $\ket{\nu_x \nu_x}$, one has $\beta_0 = \gamma_0 = \pm\cos\theta \sin\theta$ and $\{\alpha_0,\delta_0\} = \{\cos^2 \theta, \sin^2 \theta \}$ or $\{ \sin^2 \theta, \cos^2 \theta \}$ (respectively for the two distinct initial conditions). Using this condition and the expressions for $\cos2\xi$ and $\sin2\xi$ from Eq.~\eqref{xiConstraints} leads to
\begin{align}
    P_z =& \pm\cos{2\theta} {+} \frac12 \sin^2 {2\theta} \, \left[ \frac{\eta \mu_0 - \eta \mu \cos 2Q}{\sqrt{(\mu^2+ \eta^2)(\mu_0^2 + \eta^2)}}  \right], 
    \label{PzUnif} \\
    P^2 =& 1-\frac{1}{4}\sin^42\theta\bigg[\bigg(\cos2R-\frac{\eta^2+\mu_0\mu\cos2Q}{\sqrt{(\mu_0^2+\eta^2)(\mu^2+\eta^2)}}\bigg)^2
    \nonumber \\
    &+ \bigg(\sin2R-\frac{\mu\sin2Q}{\sqrt{\mu^2+\eta^2}}\bigg)^2\bigg].
    \label{PmagUnif}
\end{align}

In particular, if $\mu,\mu_0 \gg |\eta|$, then for small $\theta$ one can observe that $P_z$ oscillates with an amplitude of $\frac{\eta}{2\mu_0} \sin^2 2\theta \ll 1$, therefore oscillations are insignificant in this case. Interestingly, if we further take the limit $\eta\to0$ while $\mu,\mu_0>0$, we find that $P^2\to1$ and so the entanglement entropy $S\to0$. 
{In contrast}, for an initial state $| \nu_e \nu_x \rangle$ or $\ket{\nu_x \nu_e}$, one has $\alpha_0 = -\delta_0 = - \cos\theta \sin\theta$, and $\{ \beta_0,\gamma_0 \} = \{\cos^2\theta, -\sin^2\theta \}$ or $\{ -\sin^2\theta, \cos^2\theta \}$ (respectively for the two distinct initial conditions), which gives
\begin{align}
P_z =& {-}\frac{\pm\eta^2 \cos 2\theta + \frac12 \eta \mu_0 \sin^2 2\theta}{\sqrt{(\mu^2+ \eta^2)(\mu_0^2 + \eta^2)}} 
\nonumber \\
&{-} \cos 2Q \left[ \frac{\pm\mu\mu_0 \cos 2\theta - \frac12 \eta\mu \sin^2 2\theta}{\sqrt{(\mu^2+ \eta^2)(\mu_0^2 + \eta^2)}} \right]. 
	\label{PzMix} \\
    P^2 =& 1-\bigg[\frac{1}{2}\sin^22\theta\bigg(\cos2R-\frac{\eta^2+\mu_0\mu\cos2Q}{\sqrt{(\mu_0^2+\eta^2)(\mu^2+\eta^2)}}\bigg) 
    \nonumber \\
    &\pm\cos2\theta\bigg(\frac{\eta\mu_0-\eta\mu\cos2Q}{\sqrt{(\mu_0^2+\eta^2)(\mu^2+\eta^2)}}\bigg)\bigg]^2
    \nonumber \\
    &-\bigg[\frac{1}{2}\sin^22\theta\bigg(\sin2R-\frac{\mu\sin2Q}{\sqrt{\mu^2+\eta^2}}\bigg) +\frac{\mu\sin2Q}{\sqrt{\mu^2+\eta^2}}\bigg]^2{.}
    \label{PmagMix}
\end{align}


Here, in the limit $\mu,\mu_0 \gg |\eta|$, the oscillation amplitude {approaches} $\cos 2\theta$, which is $\sim 1$ for small $\theta$. Moreover, the instantaneous oscillation frequency is $ 2\, dQ/dt = 2 \sqrt{\mu^2 + \eta^2} \approx 2 \mu$, which indicates that the oscillations are fast (compared to the oscillations observed in the mean-field limit, which have a frequency $\sim \sqrt{|\eta| \mu}$---see, e.g., Ref.~\cite{Hannestad:2006qd}). Fast oscillations of neutrinos are also observed in certain mean-field calculations in the presence of spatial asymmetries or nonstandard neutrino-neutrino interactions~\cite{Sen:2017ogt, Dighe:2017sur}; however, in this model, the presence of many-body effects appears to render these conditions unnecessary. Fast oscillations arising from many-body effects have also been pointed out in Ref.~\cite{Rrapaj:2019pxz}, and we plan to investigate this feature in a future study. In a core-collapse supernova envelope, fast oscillations can have implications for nucleosynthesis as well as for the explosion mechanism itself. 

Additionally, we may consider the asymptotic forms of $P_z$ and $P$ for $\mu\to0$ in the cases of each initial condition. Here, our $z$-component polarization reduces to
\begin{align}
    P_z &\to \pm\cos2\theta{+}\frac{1}{2}\sin^22\theta\frac{\mathrm{sgn}[\eta]\,\mu_0}{\sqrt{\mu_0^2+\eta^2}} 
    \label{unifPzAsymp} \\
    P^2 &\to 1-\frac{1}{4}\sin^42\theta\bigg(1-2\cos2R\frac{|\eta|}{\sqrt{\mu_0^2+\eta^2}}+\frac{\eta^2}{\mu_0^2+\eta^2}\bigg)
    \label{unifP2Asymp}
\end{align}
for an initial state of $\ket{\nu_e\nu_e}$ or $\ket{\nu_x\nu_x}$. Meanwhile, from the initial states $\ket{\nu_e\nu_x}$ or $\ket{\nu_x\nu_e}$ we find 
\begin{align}
    P_z \to& {\mp}\cos2\theta\frac{|\eta|}{\sqrt{\mu_0^2+\eta^2}}{-}\frac{1}{2}\sin^22\theta\frac{\mathrm{sgn}[\eta]\,\mu_0}{\sqrt{\mu_0^2+\eta^2}} 
    \label{mixPzAsymp} \\
    P^2 \to& 1-\frac{1}{4}\sin^42\theta\bigg(1-\cos2R\frac{|\eta|}{\sqrt{\mu_0^2+\eta^2}}+\frac{\eta^2}{\mu_0^2+\eta^2}\bigg)
    \nonumber \\
    & -\cos^22\theta\frac{\mu_0^2}{\mu_0^2+\eta^2} \mp\sin^22\theta\cos2\theta\frac{\mathrm{sgn}[\eta]\,\mu_0}{\sqrt{\mu_0^2+\eta^2}}
    \nonumber \\
    & \times\bigg(\cos2R-\frac{|\eta|}{\sqrt{\mu_0^2+\eta^2}}\bigg) {~.}
    \label{mixP2Asymp}
\end{align}
Interestingly, the forms for $P_z$ depend on only the initial and the final values of $\mu$ ($\mu_0$ and $0$, in this case), while the magnitudes $P^2$ also depend on all the intermediate values of $\mu$, i.e, $0<\mu<\mu_0$, through terms proportional to $\cos2R$. If we also take $\mu_0\gg|\eta|$, this additional dependence vanishes for the {\color{black} monoflavor initial states (i.e., $\ket{\nu_e\nu_e}$ and $\ket{\nu_x\nu_x}$) but not for the mixed-flavor initial states}.

{Finally we note that, for a two-neutrino system, the results obtained for $P^2$ must be independent of our choice of particle to take the partial trace over; $P$ shares a one-to-one relationship with entanglement entropy $S$, as per Eq.~\eqref{concEntropy}, which for a bipartite pure-state system does not depend on which of the two partitions is being traced over. However, this invariance is not shared by $P_z$, which will differ between the two neutrinos' subsystems. To explicitly see this fact, one can show that the first particle's $z$ component of polarization is given by $P_z=|\alpha|^2+|\beta|^2-|\gamma|^2-|\delta|^2$, in contrast with the form for the second particle in Eq.~\eqref{Pz2general}.}

\subsection{An example of the mean-field limit}\label{sec:N=2mft}

To illustrate the mean-field limit of the Hamiltonian in Eq.~(\ref{singleH}), here we show a simple example with two neutrinos. It is instructive to subtract constant terms {proportional to} $\vec{J}_1^2$ and $\vec{J}_2^2$, to obtain
\begin{equation}
	H = -\omega_1 J_1^z -\omega_2 J_2^z + 2 \mu\vec{J}_1\cdot\vec{J}_2 .
	\label{singleH2}
\end{equation}
{Expressing the} average of an operator $X$ over the mean field as $\langle X \rangle_M$, we can write the quadratic term in the above equation as 
\begin{equation}
\vec{J}_1\cdot\vec{J}_2 \approx  \vec{J}_1\cdot \langle \vec{J}_2 \rangle_M + \langle \vec{J}_1 \rangle_M  \cdot\vec{J}_2 - \langle  \vec{J}_1 \rangle_M \cdot 
\langle \vec{J}_2 \rangle_M {~,}
\end{equation}
where one observes that the Hamiltonian in Eq. (\ref{singleH}) 
becomes, up to a constant term,  a direct sum of Hamiltonians describing individual neutrinos 
\begin{eqnarray}
	H &=& \left( -\omega_1J_1^z + 2 \mu\vec{J}_1\cdot \langle \vec{J}_2 \rangle_M \right)  + \left( - \omega_2J_2^z + 2 \mu\vec{J}_2\cdot \langle \vec{J}_1 \rangle_M \right) \nonumber \\
	&-&  2 \mu \langle \vec{J}_1 \rangle_M \cdot \langle \vec{J}_2 \rangle_M ~.
	\label{singleH3}
\end{eqnarray}
With this approximation, the Hamiltonian can be written as a direct sum of the individual one-body Hamiltonians. Consequently, the eigenstates are direct products of the single neutrino states, and therefore unentangled. The time evolution of these operators takes the form  
\begin{equation}
\label{J1evo}
 \frac{\partial J^i_1}{\partial t} =  \varepsilon_{ijk} \left( \omega_1 B^j + 
 2 \mu \langle J^j_2 \rangle_M \right) J^k_1
\end{equation}
and a similar equation for $\vec{J}_2$, where the indices $i,j,k$ represent the Cartesian components.  
Note that a consistent mean field requires that Eq. (\ref{J1evo}) imply
\begin{equation}
 \frac{\partial \langle J^i_1 \rangle}{\partial t} =  \varepsilon_{ijk} \left( \omega_1 B^j + 
 2 \mu \langle J^j_2 \rangle \right) \langle J^k_1 \rangle {~,}
\end{equation}
and a similar equation for $\vec{J}_2$. Defining the mean-field polarization vectors as $\vec{P}_1=2 \langle \vec{J}_1 \rangle_M$, $\vec{P}_2= 2 \langle \vec{J}_2 \rangle_M$, and $\vec{P} = \vec{P}_1 + \vec{P}_2$, we obtain 
\begin{equation}
\frac{\partial \vec{P}_1}{\partial t} =  \left( \omega_1 \vec{B} + \mu \vec{P} \right) \times \vec{P}_1
\end{equation}
and 
\begin{equation}
\frac{\partial \vec{P}_2}{\partial t} =  \left( \omega_2 \vec{B} + \mu \vec{P} \right) \times \vec{P}_2.
\end{equation}
From these equations we see that both $|\vec{P}_1|$ and $|\vec{P}_2|$ are constants in time (and equal to one for this choice of normalization), as expected for unentangled particles. Additionally, by adding the two equations, it is easy to see that $\vec{B}\cdot\vec{P}$ is also conserved, which is consistent with the commutation of total $J^z$ with the Hamiltonian in Eq.~\eqref{singleH}.

\begin{figure*}[htb]
\begin{center}
    \subfloat[\label{Pzvsr_ee}]{
	    \includegraphics[width=0.49\textwidth]{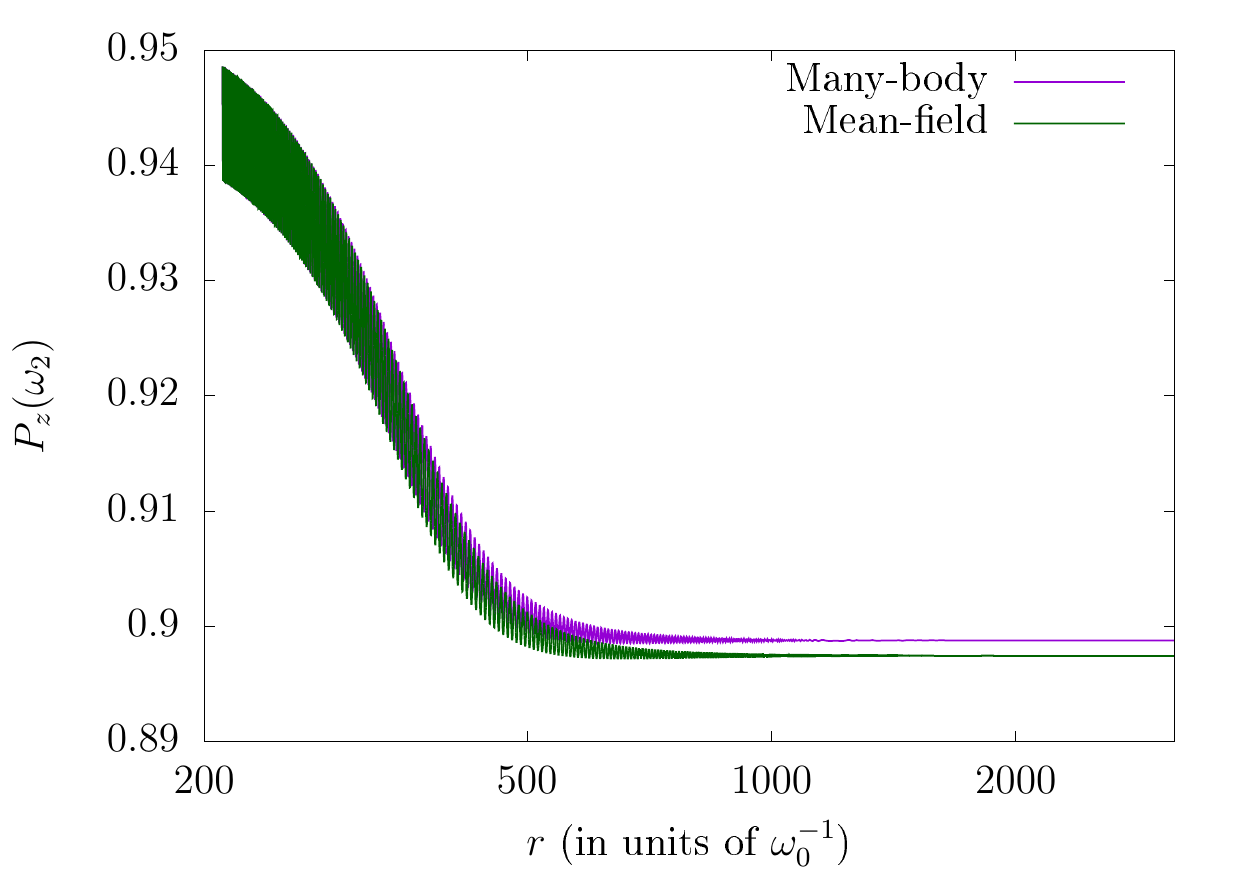}}
	~
	\subfloat[\label{ent_ee}]{
	    \includegraphics[width=0.49\textwidth]{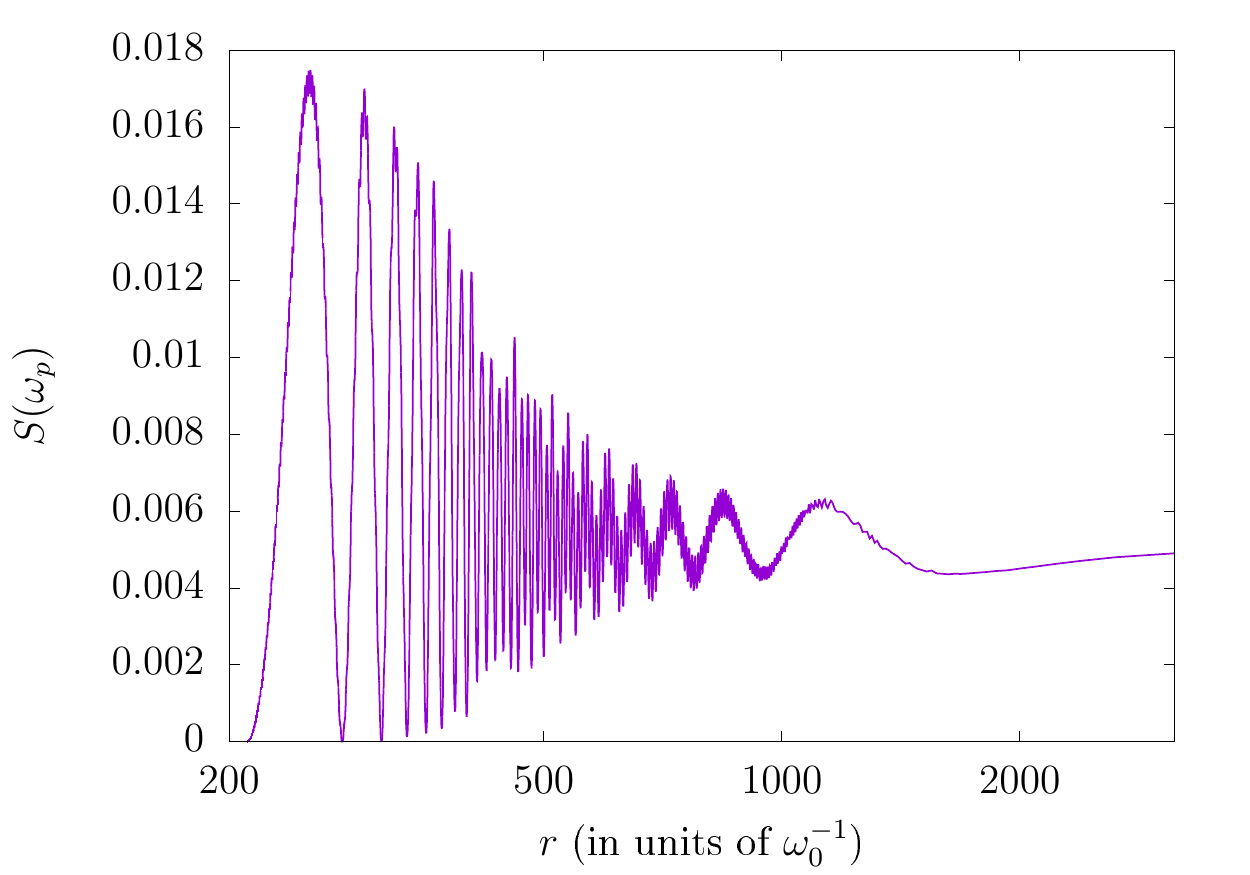}}
\end{center}
	\caption{Left: Evolution of $P_z$ of the neutrino with frequency $\omega_2$ as a function of radius $r$, for a two-neutrino system with initial condition $\ket{\nu_e\nu_e}$. The two lines represent the adiabatic evolution using the many-body treatment and the mean-field evolution, respectively. Right: Entanglement entropy as a function of $r$, for the same system. For an $N=2$ system, the entanglement entropy of both neutrinos must be identical 
	by Eq.~\eqref{EntEntropy}.
}
	\label{fig:Pzevol_ee}
\end{figure*}

\begin{figure*}[htb]
\begin{center}
    \subfloat[\label{Pzvsr_emu}]{
	    \includegraphics[width=0.49\textwidth]{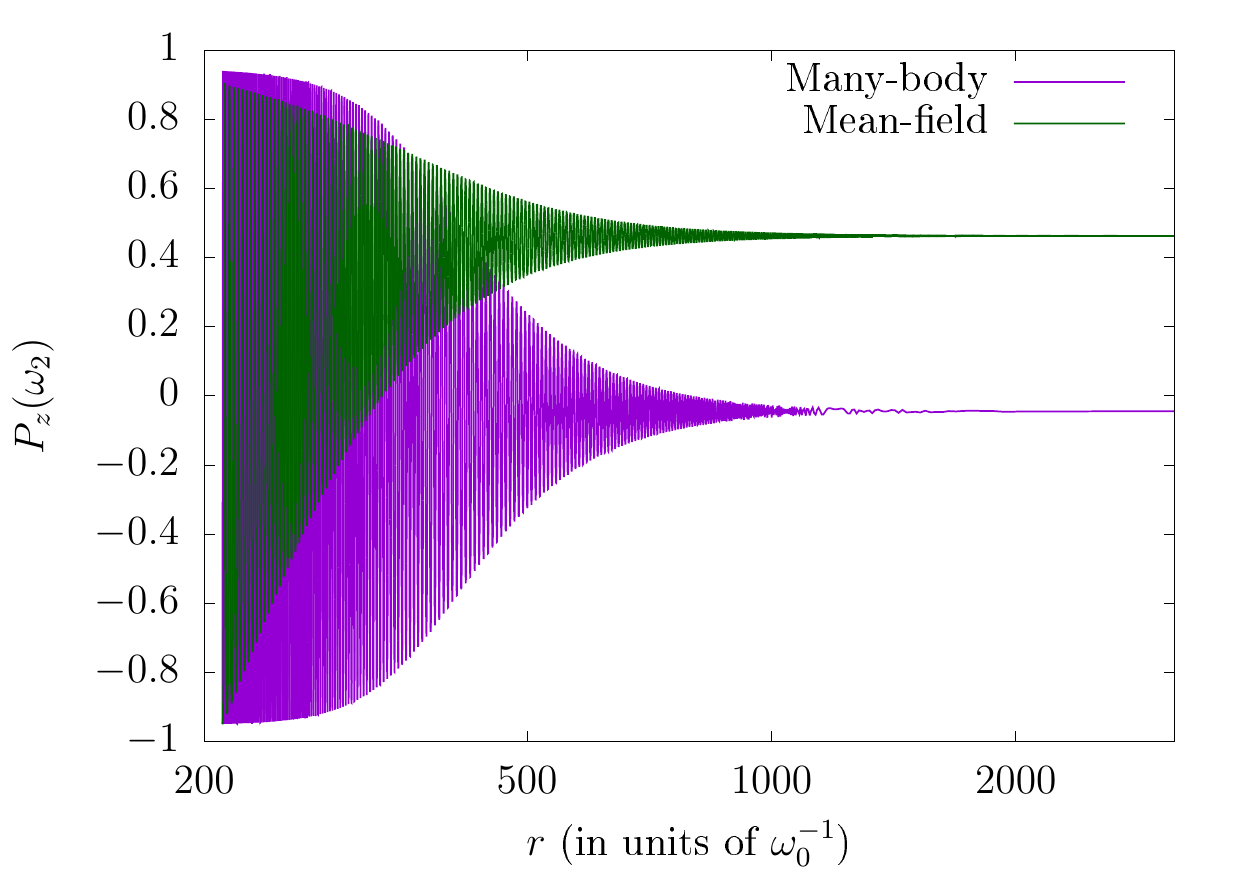}}
	~
	\subfloat[\label{ent_emu}]{
	    \includegraphics[width=0.49\textwidth]{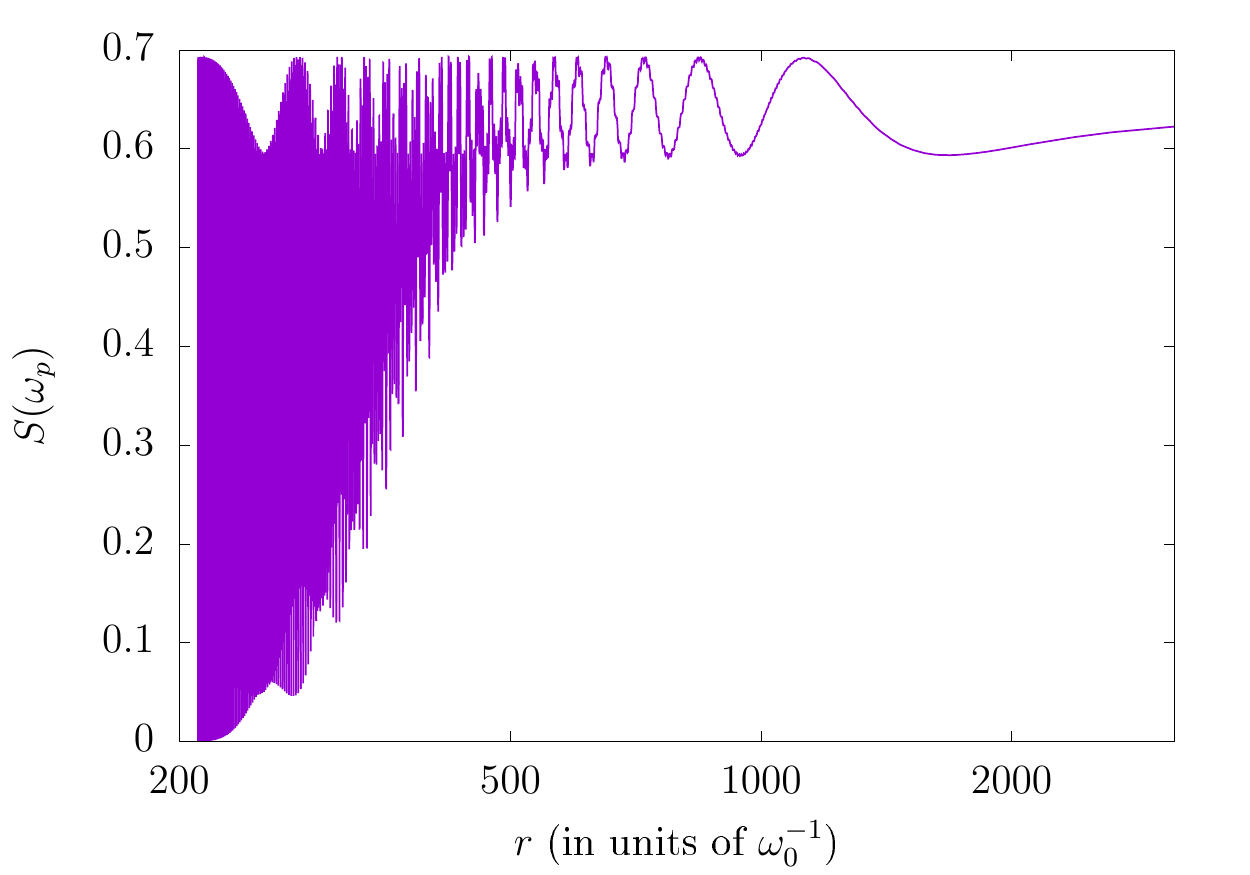}}
\end{center}
	\caption{Same as Fig.~\ref{fig:Pzevol_ee}, but for a system starting with an initial condition $\ket{\nu_e\nu_x}$.}
	\label{fig:Pzevol_emu}
\end{figure*}

In Figs.~\ref{fig:Pzevol_ee} and \ref{fig:Pzevol_emu}, we show the evolution of $P_z$ of neutrino 2 as a function of the radius, for an $N=2$ system with initial conditions $\ket{\nu_e\nu_e}$ and $\ket{\nu_e\nu_x}$, respectively. In addition, we also show the dependence of entropy on $\mu$ for each of the systems, alongside the corresponding $P_z$ evolution plots. Here, $\mu$ is taken to vary with the radius in accordance with the single-angle bulb model expression, given in Eq.~\eqref{singleMu}. A comparison of the two figures reveals that for the system with a $\ket{\nu_e\nu_x}$ initial condition, the entanglement entropy is quite high, and this is associated with the evolution of $P_z$ differing substantially in the many-body vs mean-field treatments. In contrast, for the system starting from a $\ket{\nu_e\nu_e}$ configuration, the entropy never grows too large, {and the mean-field evolution does not deviate much from that of the many-body solution}.

\section{Numerical results for more than two neutrinos} \label{section:N>2}

In this section we 
quantify the entanglement between neutrinos in time-evolved many-body states described in Section~\ref{section:Evolution}. 
For this purpose, we choose systems with different initial configurations, wherein each neutrino is either in the $\nu_e$ or in the $\nu_x$ flavor, and we apply the methods described in Sec.~\ref{section:GenEntangle} to calculate measures of entanglement for these systems.

To determine this entanglement entropy at each $\mu<\mu_0$ (or equivalently $r>r_0$), we first construct a density matrix from our evolved state, in the mass basis.\footnote{One can just as well carry out this process using the evolved state in the flavor basis. The only change to our results will be in the orientation of the Bloch vector between the two bases, for $\theta\neq0$.} For calculating entanglement between the neutrino with the highest value of $\omega$
and the other $N-1$ neutrinos, we compute a reduced density matrix by taking a partial trace over the subspaces of the $N-1$ other neutrinos, as per Eq.~\eqref{densityRed}. With this effective one-body density matrix we can calculate the entropy of entanglement between this neutrino and the rest. Results found from this procedure are displayed in Fig.~\ref{GeneralEntropyR}. As we expect, $S\sim0$ in the limit of small $r\sim R_\nu$. Further, as $r$ grows ($\mu$ decreases), we find that entropy values eventually level off to constant values as the collective oscillation strength becomes much smaller than vacuum oscillation frequencies (i.e., $\mu\ll\omega_0$).
For the results displayed in this paper, we use typical values for the parameters mentioned thus far, summarized in Table \ref{parameters}.



{We investigated other measures of entanglement for these evolved states in addition to the entanglement entropy. For example, we calculated logarithmic negativity of the subsystem for the neutrino with vacuum oscillation frequency $\omega_N$, as prescribed in Ref.~\cite{Vidal:2002zz}.
Additionally,} we calculated the entanglement of formation, as defined by Ref~\cite{wootters1998entanglement}, of the effective mixed, two-neutrino state corresponding with oscillation frequencies $\omega_{N-1}$ and $\omega_N$ after taking the partial trace over all $N-2$ other neutrinos. We verified that both measures yield qualitatively similar behavior to that of entanglement entropy and are therefore not presented in this article.


{Moreover, w}e seek to determine whether many-body effects can drastically alter the evolution of a multineutrino system, as compared with the evolution observed in the mean-field limit. To this end, we plot the evolution of $P_z$ {of each neutrino} in both the many-body and the mean-field picture.

\begin{figure*}[htb]
\begin{center}
    \subfloat[\label{Pzvsr8_001}]{
	    \includegraphics[width=0.49\textwidth]{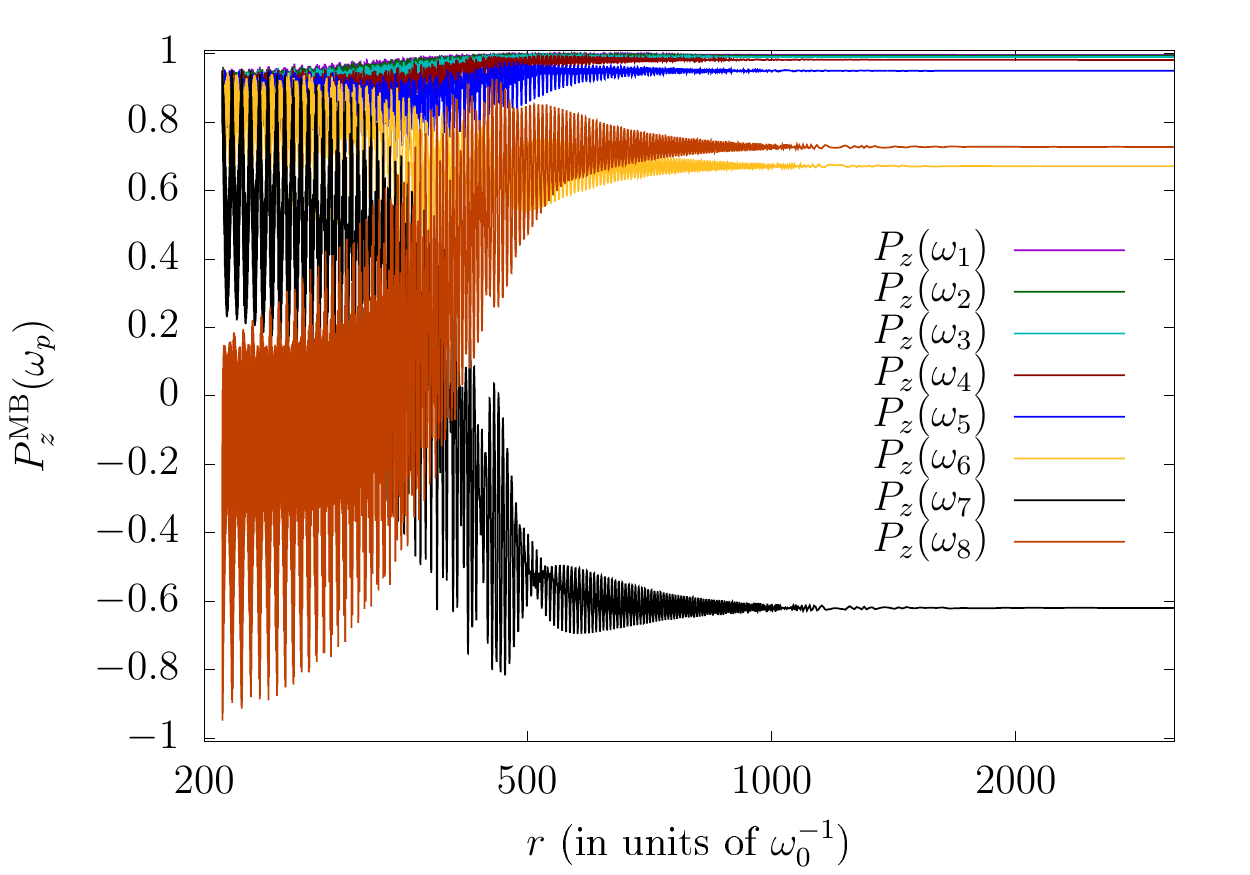}}
	~
	\subfloat[\label{Pzvsr8_001_MF}]{
	    \includegraphics[width=0.49\textwidth]{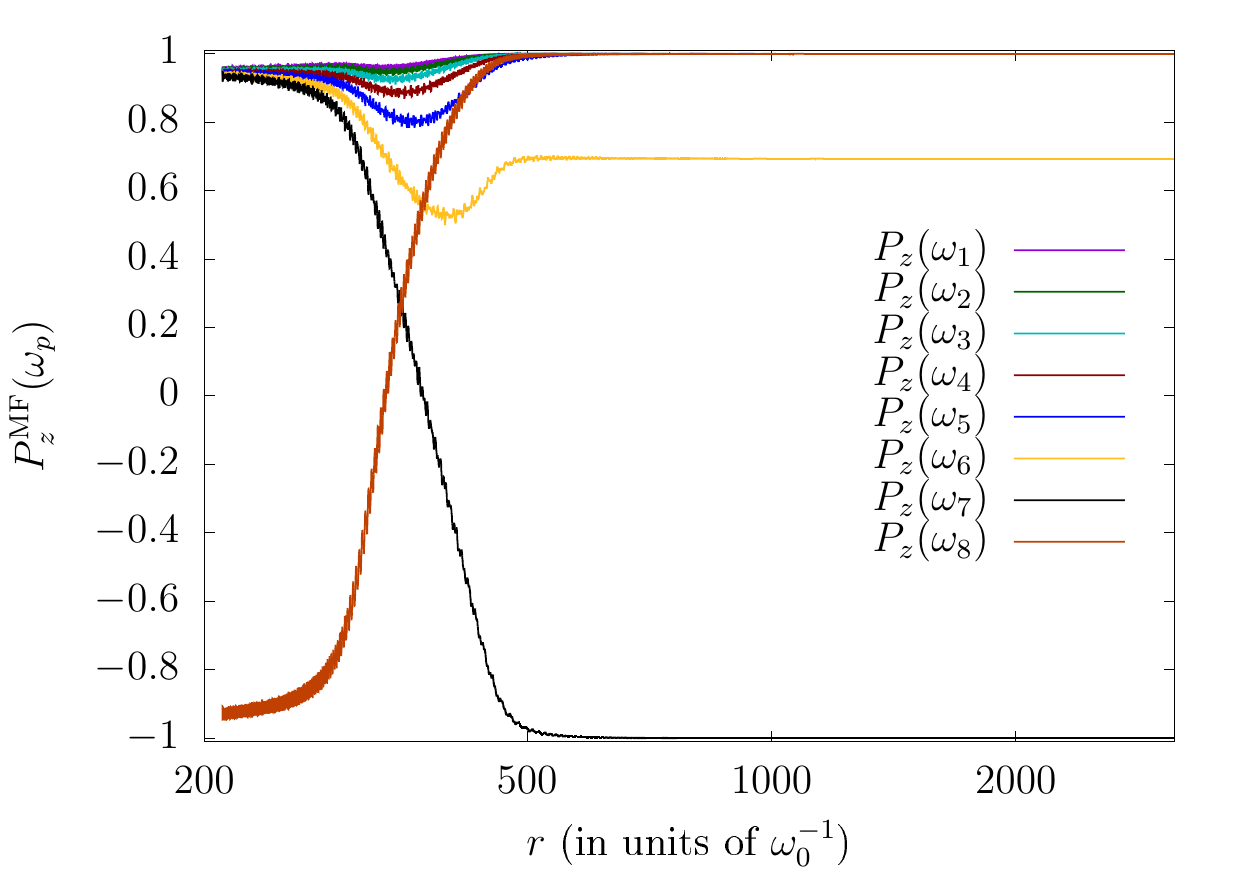}}
\end{center}
	\caption{Evolution of $P_z$ of each neutrino as a function of radius $r$, for an $N=8$ system with an initial configuration consisting of a $\nu_e$ at each of the frequencies $\omega_1,\ldots,\omega_7$ and a $\nu_x$ at $\omega_8$. Left: Evolution of the system calculated in the adiabatic many-body framework. Right: Evolution calculated using the mean-field approximation.}
	\label{fig:Pzvsr8_001}
\end{figure*}

\begin{figure*}[htb]
\begin{center}
    \subfloat[\label{Pzvsr8_015}]{
	    \includegraphics[width=0.49\textwidth]{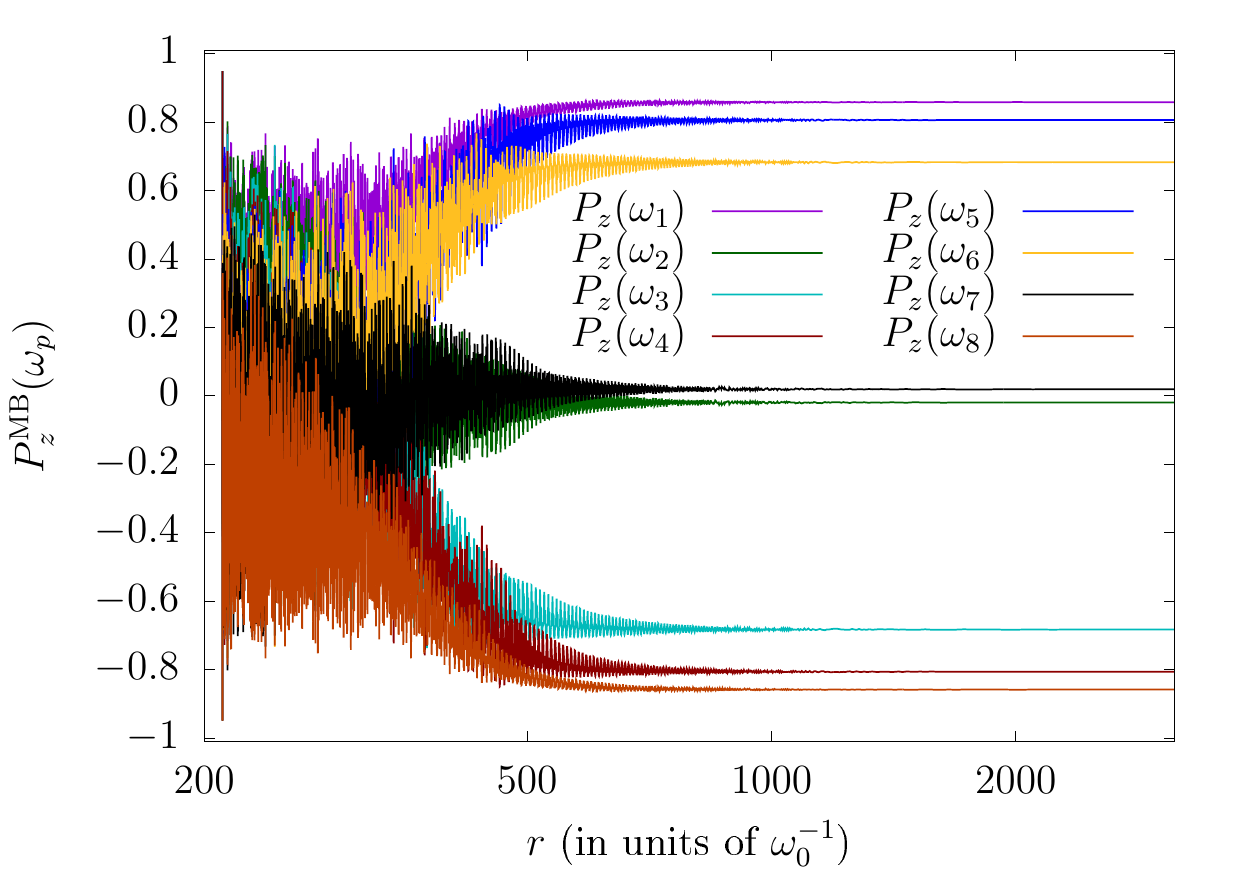}}
	~
	\subfloat[\label{Pzvsr8_015_MF}]{
	    \includegraphics[width=0.49\textwidth]{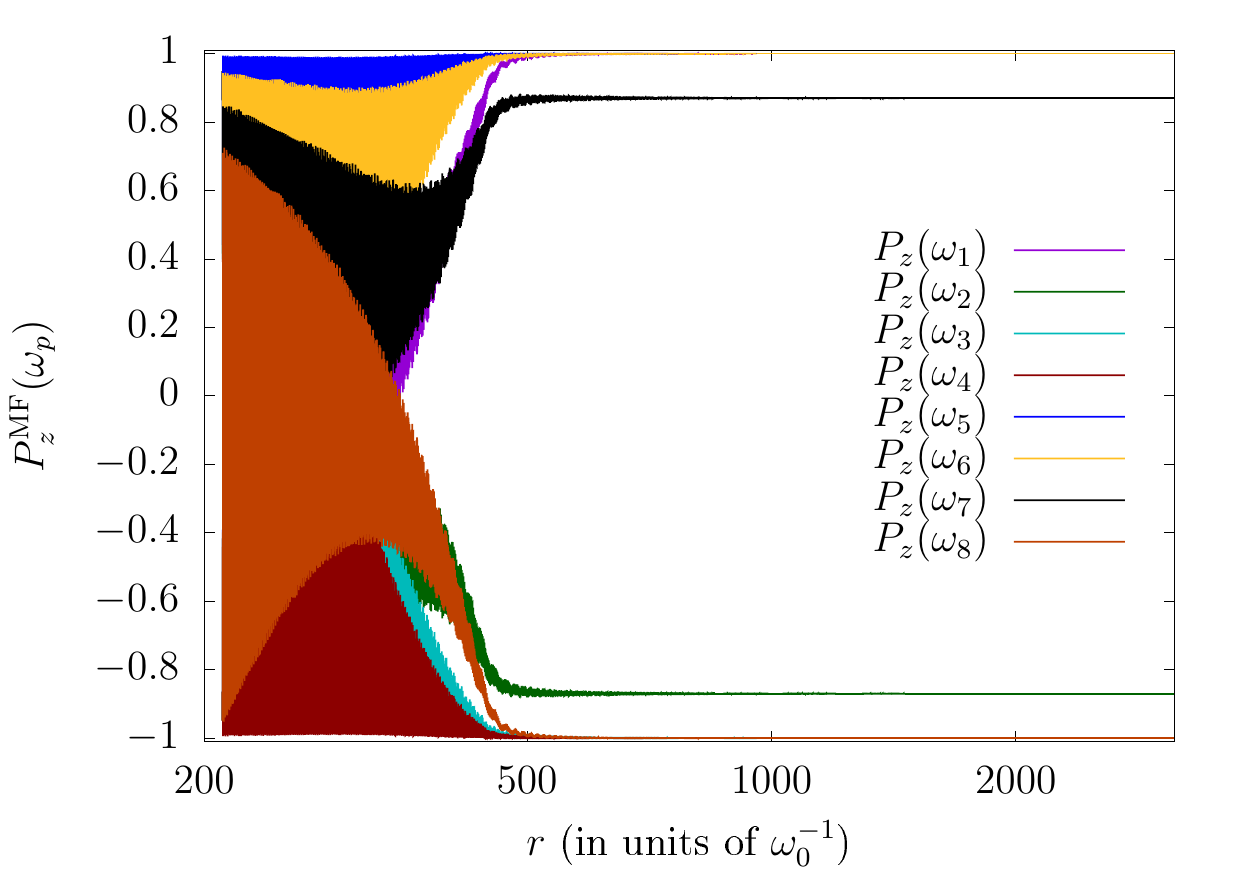}}
\end{center}
	\caption{
	{Same as Fig.~\ref{fig:Pzvsr8_001}, but for a system with an initial configuration consisting of a $\nu_e$ at each of the frequencies $\omega_1,\ldots,\omega_4$, and a $\nu_x$ at each of $\omega_5,\ldots,\omega_8$.}
	}
	\label{fig:Pzvsr8_015}
\end{figure*}

As described in Sec.~\ref{section:N=2}, one can obtain analytic closed-form expressions for the evolving quantities for an $N=2$ system, which may then be used for comparison with the corresponding mean-field solution. Such a comparison is presented in Figs.~\ref{fig:Pzevol_ee} and \ref{fig:Pzevol_emu}. We can repeat this comparison for systems with larger $N$ using numerical methods as outlined in Sec.~\ref{section:Evolution} and Appendix~\ref{Appendix_Eigen}. Shown in Figs.~\ref{fig:Pzvsr8_001} and \ref{fig:Pzvsr8_015} are $P_z$ (in both many-body and mean-field calculations) and $S$ results of each neutrino for a system with $N=8$, with two different initial states. In Fig.~\ref{fig:Pzvsr8_001}, we depict a system starting from an initial state with neutrinos with frequencies $\omega_1,\ldots,\omega_7$ in the $\nu_e$ flavor, and the $\omega_8$ neutrino in the $\nu_x$ flavor, whereas in Fig.~\ref{fig:Pzvsr8_015}, we pick an initial state consisting of $\nu_e$ at frequencies $\omega_1,\ldots,\omega_4$, and $\nu_x$ at $\omega_5,\ldots,\omega_8$. Furthermore, we illustrate in Figs.~\ref{fig:spec&ent_001} and \ref{fig:spec&ent_015} how entanglement entropy can reflect the discrepancies between many-body and mean-field predictions of time-evolved mass-basis spectra of the ensemble, for the same initial states as in Figs.~\ref{fig:Pzvsr8_001} and~\ref{fig:Pzvsr8_015} respectively. In the final spectra plots comparing mean-field and many-body predictions, Figs.~\ref{Pzspec_001} and~\ref{Pzspec_015}, we observe a weaker spectral swap when including many-body effects, as asymptotic values of $P_z$ at the frequencies surrounding the swap do not approach $\pm 1$, unlike their counterparts in the mean-field theory. Further, as previously outlined in Sec.~\ref{section:OverResults}, there appears to be a correlation, illustrated in Fig.~\ref{fig:PvS}, between the entanglement entropy of each neutrino with the ensemble and the deviation of its polarization vector $z$ component, $P_z$, relative to the mean-field limit.

\begin{figure*}[htb]
\begin{center}
    \subfloat[\label{Pzspec_001}]{
	    \includegraphics[width=0.49\textwidth]{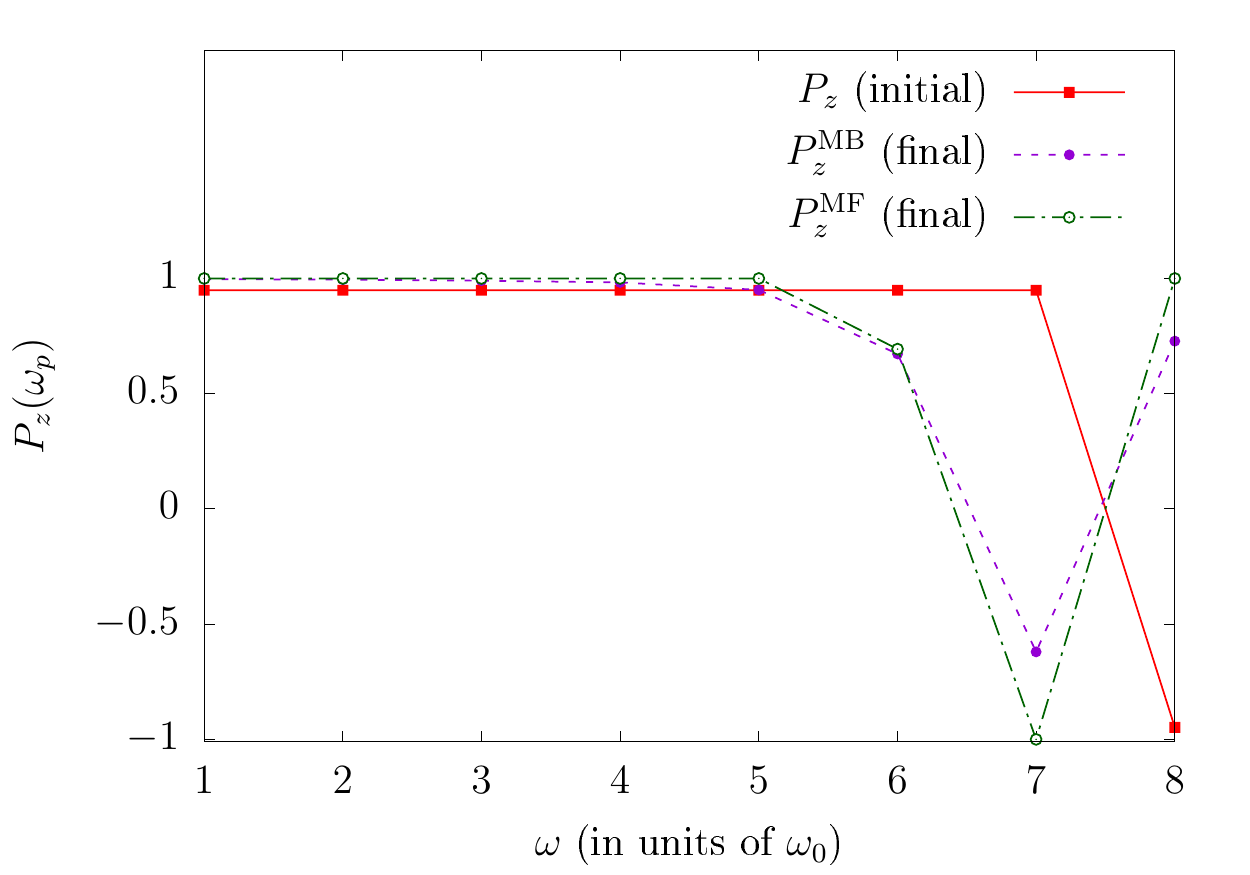}}
	~
	\subfloat[\label{ent_001}]{
	    \includegraphics[width=0.49\textwidth]{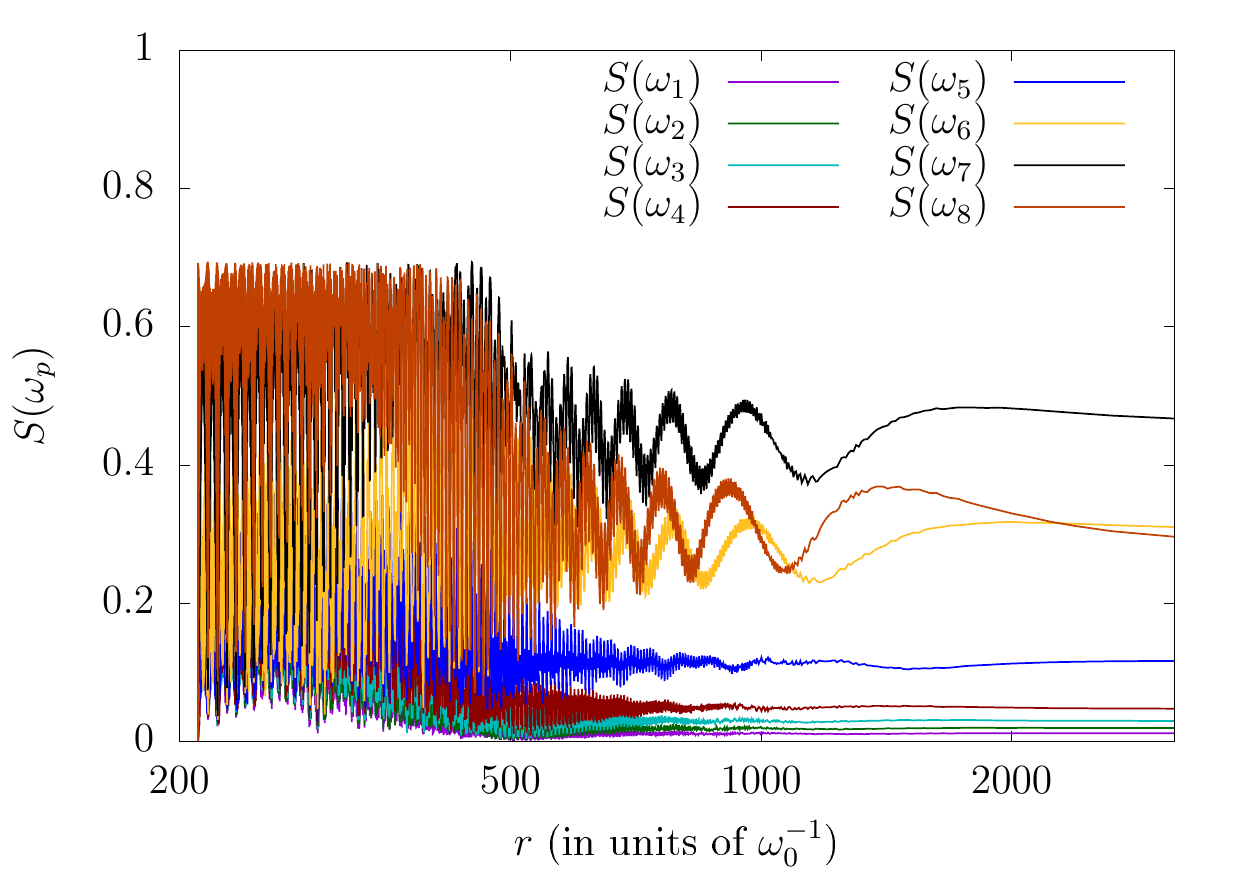}}
\end{center}
	\caption{Left: Comparison of $P_z$ spectra for the mean-field (MF) and many-body (MB) calculations, for an $N=8$ system with initial configuration consisting of a $\nu_e$ at each of the frequencies $\omega_1,\ldots,\omega_7$ and a $\nu_x$ at $\omega_8$. Right: Evolution of the entanglement entropies, $S(\omega_p)$, of each neutrino with the rest of the ensemble.}
	\label{fig:spec&ent_001}
\end{figure*}

\begin{figure*}[htb]
\begin{center}
    \subfloat[\label{Pzspec_015}]{
	    \includegraphics[width=0.49\textwidth]{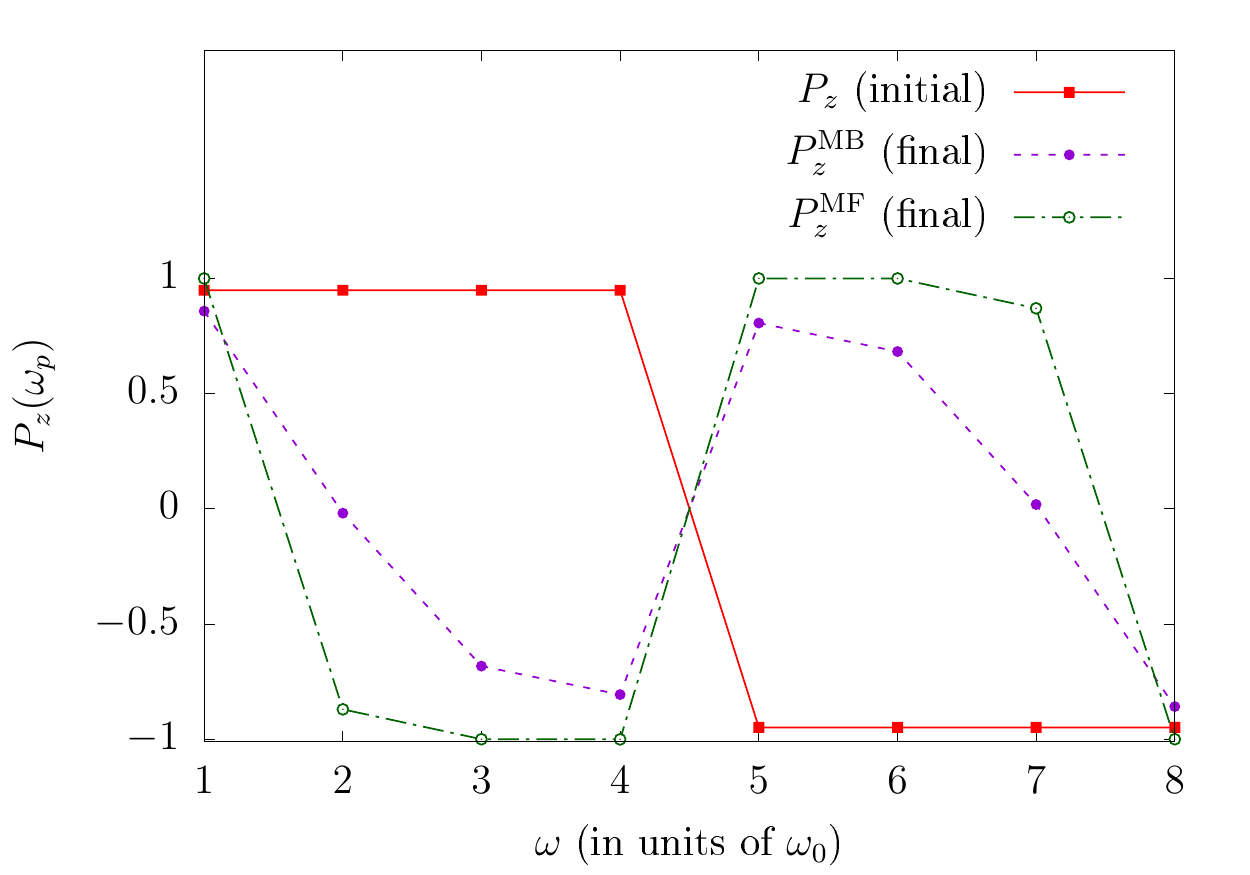}}
	~
	\subfloat[\label{ent_015}]{
	    \includegraphics[width=0.49\textwidth]{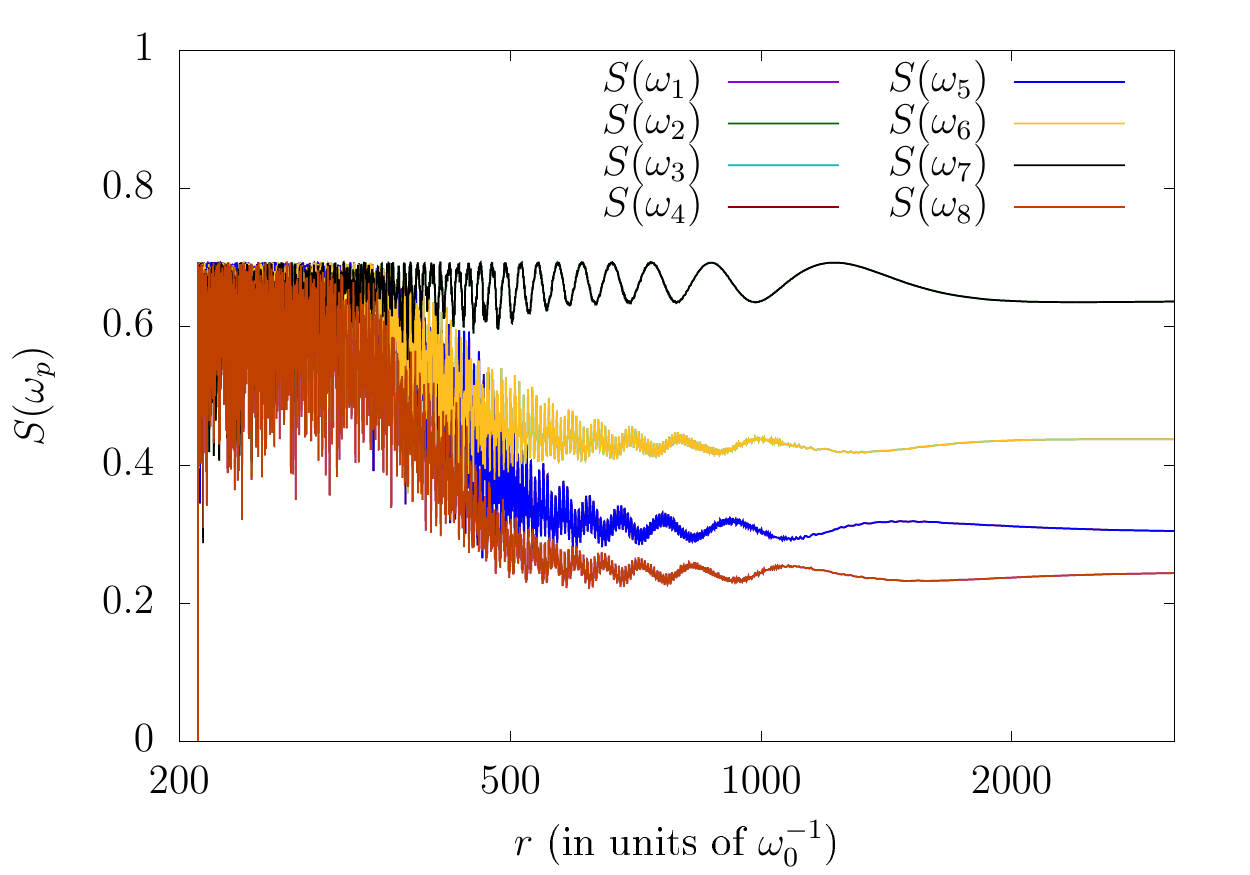}}
\end{center}
	\caption{
	{Same as Fig.~\ref{fig:spec&ent_001}, but for a system with an initial configuration consisting of a $\nu_e$ at each of the frequencies $\omega_1,\ldots,\omega_4$, and a $\nu_x$ at each of $\omega_5,\ldots,\omega_8$.}
	}
	\label{fig:spec&ent_015}
\end{figure*}

\section{Conclusions} \label{section:Conclusion}

Collective neutrino oscillations represent a nonlinear quantum problem with a significant impact on astrophysics and cosmology. However, calculating collective oscillations of a very large number of neutrinos present in these settings is a challenging many-body problem. To simplify the problem one often resorts to the mean-field approximation albeit at the cost of losing entanglement between individual neutrinos. Hence, measures of entanglement could be used to quantify the deviation of the mean-field approximation from many-body results. We showed  that the entanglement entropy, which is zero within the mean-field approximation but nonzero in the many-body calculations, is a useful quantity for identifying regimes where mean-field theory may be insufficient. 

In this paper, we examined collective neutrino oscillations for systems with small numbers of neutrinos, for which we can easily obtain  many-body solutions using the methods described in Ref.~\cite{Patwardhan:2019zta}, and we showed that the mean-field approximation can be incomplete. We also demonstrated that, for systems consisting initially of a single flavor, the third component of the polarization vector for the neutrino with the highest frequency deviates more and more from the mean-field value as the number of neutrinos is increased from $N=2$ to $9$. Correspondingly the asymptotic value of the entanglement entropy of those neutrinos  increases as the number of neutrinos increases. 

Our results exhibit a strong dependence on the initial conditions. We also examine various initial conditions, consisting of both $\nu_e$ and $\nu_x$ flavors, where the evolution in the mean field results in a spectral swap. In the many-body calculations, we observe that the swap is largely preserved, but appears weaker because of the $P_z$ values surrounding the swap frequency being further from $\pm 1$ compared to the mean-field case. This deviation is also correlated with the entanglement entropy of those neutrinos being large.

Calculations performed using the mean-field approximation have revealed a lot of interesting physics about collective behavior of neutrinos in astrophysical environments. Here we have explored possible scenarios where further interesting features can arise by going beyond this approximation.


\begin{acknowledgments}
We thank Y.~Pehlivan, E.~Rrapaj, and P.~Claeys for helpful conversations. This work was supported in part by the U.S. Department of Energy, Office of Science, Office of High Energy Physics, under Award No.~DE-SC0019465. 
It was also supported in part by the U.S. National Science Foundation
Grants No.~PHY-1630782 and No.~PHY-1806368.
\end{acknowledgments}

\appendix

\section{Determination of Eigenvalues and Eigenstates} \label{Appendix_Eigen}

In this section, we outline how to determine the eigenvalues and eigenstates of the Hamiltonian and the conserved charge operators described in Sec.~\ref{sec:model}. A more detailed exposition can be found in our previous study, Ref.~\cite{Patwardhan:2019zta}---however, here we point out, at the end of our summary, a symmetry in the eigenvalue equations that can be used to reduce the computational time. Though arguments in this appendix {can be generalized to the higher dimensional (i.e., $j_p>1/2$) representations}, in this discussion we mainly focus on the case of $j_p=1/2$ for all $p$, which corresponds to having a single neutrino at each $\omega_p$. 

Because the operator $J^z$ (the $z$ component of the total isospin) commutes with the Hamiltonian, each energy eigenstate will have a definite eigenvalue $m=-N/2,-N/2+1,\ldots,+N/2$ of $J^z$. For a given $N$ and $m$ it can be shown that there are $^N C_{N/2-m}$ energy states with total $z$-component isospin number $m$. Since this quantity is conserved even in the event of degeneracies in energy eigenvalues~\cite{Birol:2018qhx}, it is helpful to categorize solutions to our system by their corresponding value of $m$. {Additionally, the solutions for eigenvalues of conserved charges $h_p$ allow us to easily determine the $m$ value of a given state}.

It is possible to obtain an algebraic system of coupled equations for the eigenvalues $\epsilon_p$ of the charge operators $h_p$, for any $j_p$. For $j_p=1/2$ we find there are $N$ coupled quadratic equations constraining these values \cite{Dimo:2018tdu}:
\begin{equation}
	\epsilon_p^2 = \mu\sum_{\substack{q=1\\q\neq p}}^N \frac{\epsilon_q}{\omega_p-\omega_q}+\frac{1}{4}+\frac{3}{4}\mu^2\sum_{\substack{q=1\\q\neq p}}^N\frac{1}{(\omega_p-\omega_q)^2}{~.}
	\label{quadraticEpsilons}
\end{equation}
The eigenvalues $\epsilon_p$ can then be used to obtain the energy eigenvalues of the Hamiltonian: 
\begin{equation}
	E=\sum_{p=1}^N\omega_p\epsilon_p+\frac{3}{4}\mu^2N {~.}
	\label{EnergyEpsilon}
\end{equation}

Further, these charge eigenvalues are related to another system of $N$ parameters $\Lambda_p$~\cite{Balantekin:2018mpq}:
\begin{equation}
	\epsilon_p = \frac{1}{2}\mu\sum_{\substack{q=1\\q\neq p}}^M\frac{1}{\omega_p-\omega_q}\pm\frac{1}{2}-\mu \Lambda_p ~{.}
	\label{epsLam}
\end{equation}
These parameters similarly satisfy a system of $N$ coupled quadratic equations for $j_p=1/2$ \cite{Babelon_2007,Faribault:2011rv}:
\begin{equation}
	F_p(\vec{\Lambda})\equiv\Lambda_p^2\mp\frac{1}{\mu}\Lambda_p-\mu\sum_{\substack{q=1\\q\neq p}}^N\frac{\Lambda_p-\Lambda_q}{\omega_p-\omega_q}=0 {~,}
	\label{quadraticLambdas}
\end{equation}
where $\vec{\Lambda}\equiv(\Lambda_1,\ldots,\Lambda_N)$. The {choice of} alternative {sign} in Eqs.~\eqref{epsLam} and \eqref{quadraticLambdas} is determined by the choice of formalism (i.e., raising or lowering, respectively) for defining these parameters as per the Bethe ansatz equations \cite{Claeys:2017sp, Claeys:2018zwo, Birol:2018qhx, Patwardhan:2019zta}. With these $N$ parameters 
{one is} equipped to determine the overlaps (up to normalization) of all energy eigenstates $\ket{q}$ with the mass product basis elements $\ket{\nu_{i_1},\ldots,\nu_{i_N}}$:
\begin{equation}
	\braket{\nu_{i_1},\ldots,\nu_{i_N}|q} = \det(\mathcal{J}(\vec{\Lambda}))
\end{equation}
where $\mathcal{J}$ is a $\kappa \times \kappa$ matrix, with $\kappa = N/2 \pm m$, defined by
\begin{equation}
	\mathcal{J}_{\alpha\beta} \equiv \begin{cases} 
	\Lambda_{p_\alpha} - \sum_{\substack{\gamma=1\\\gamma\neq \alpha}}^{N/2\pm m} \frac{1}{\omega_{p_\alpha}-\omega_{p_\gamma}} & \mathrm{if}\: \alpha=\beta \\ 
	\phantom{\Lambda_{p_\alpha} }-\frac{1}{\omega_{p_\alpha}-\omega_{p_\beta}} & \mathrm{if}\: \alpha\neq\beta 
	\end{cases}
	\label{Jmatrix}
\end{equation}
with indices $p_\alpha$ only spanning values satisfying $i_{p_\alpha}=1$ ($2$) \cite{Claeys:2017sp, Claeys:2018zwo}. That is, this matrix only considers the frequency bins $\omega_{p_\alpha}$ occupied by a neutrino with isospin-up (-down) 
in the state $\ket{\nu_{i_1},\ldots,\nu_{i_N}}$. Moreover, elements of this matrix $\mathcal{J}$ are evaluated at the particular solution $\vec{\Lambda}$ to Eqs.~\eqref{quadraticLambdas} that satisfies the condition { $\pm\mu\Lambda_{j}\to q_j$} as $\mu\to0$ for $j=1,\ldots,N$, where $q=\sum_{j=1}^N(q_j-1)\,2^{N-j}$ is the binary representation of $q$. Alternatively, the eigenstates (again, up to a normalization factor) can be constructed from the identities
\begin{equation}
	\ket{q} = \frac{1}{\kappa!}\sum_{\sigma\in \mathrm{Sym}(\kappa)}\mathrm{sgn}(\sigma)\mathrm{tr}_\sigma(A)\ket{\nu_{y}\ldots\nu_y},
	\label{estatesFromMatrix}
\end{equation}
where $y=2$ ($1$) in the raising (lowering) formalism, $\mathrm{Sym}(\kappa)$ is the symmetry group of $\kappa$ letters, $\mathrm{Tr}_\sigma(A)=\mathrm{Tr}(A^{f_1})\cdots\mathrm{Tr}(A^{f_n})$ for a permutation $\sigma$ of cycle type $(f_1,\ldots,f_n)$, and the individual traces are 
\begin{equation}
	\mathrm{Tr}(A^f) = \sum_{p_1=1}^N\cdots\sum_{p_f=1}^NJ_{p_1}^{\pm}\cdots J_{p_f}^{\pm}\sum_{m=1}^f\Lambda_{p_m}\prod_{\substack{l=1\\l\neq m}}^f\frac{1}{\omega_{p_l}-\omega_{p_m}}.
	\label{estateFactors}
\end{equation}
Here, $A$ is a diagonal matrix of Gaudin raising (lowering) operators, as defined in Ref. \cite{Patwardhan:2019zta}. One can determine the corresponding $m$ for a solution using either $\vec{\Lambda}$ or $\vec{\epsilon}$, as $\sum_{p}\mu\Lambda_p=\pm\kappa$ and $\sum_{p}\epsilon_p=-m$. For this reason as well, values of $m$ are convenient for categorizing solutions. 

Note that from a solution $\vec{\Lambda}$ obtained in a raising (lowering) formalism one can obtain another solution in the lowering (raising) formalism, given by $\vec{\Lambda}\mp(1/\mu,\ldots,1/\mu)$ \cite{Claeys:2017sp,Patwardhan:2019zta}. Furthermore, by selecting either the raising or the lowering formalism for determining an eigenstate so that $\kappa=\min\{N/2+m,N/2-m\}$, considerable computation time may be reduced in calculating these states in the mass basis.

\begin{figure*}[htbp]
\begin{center}
	\subfloat[$m=3/2$\label{EnergyEnt1}]{		\includegraphics[width=0.49\textwidth]{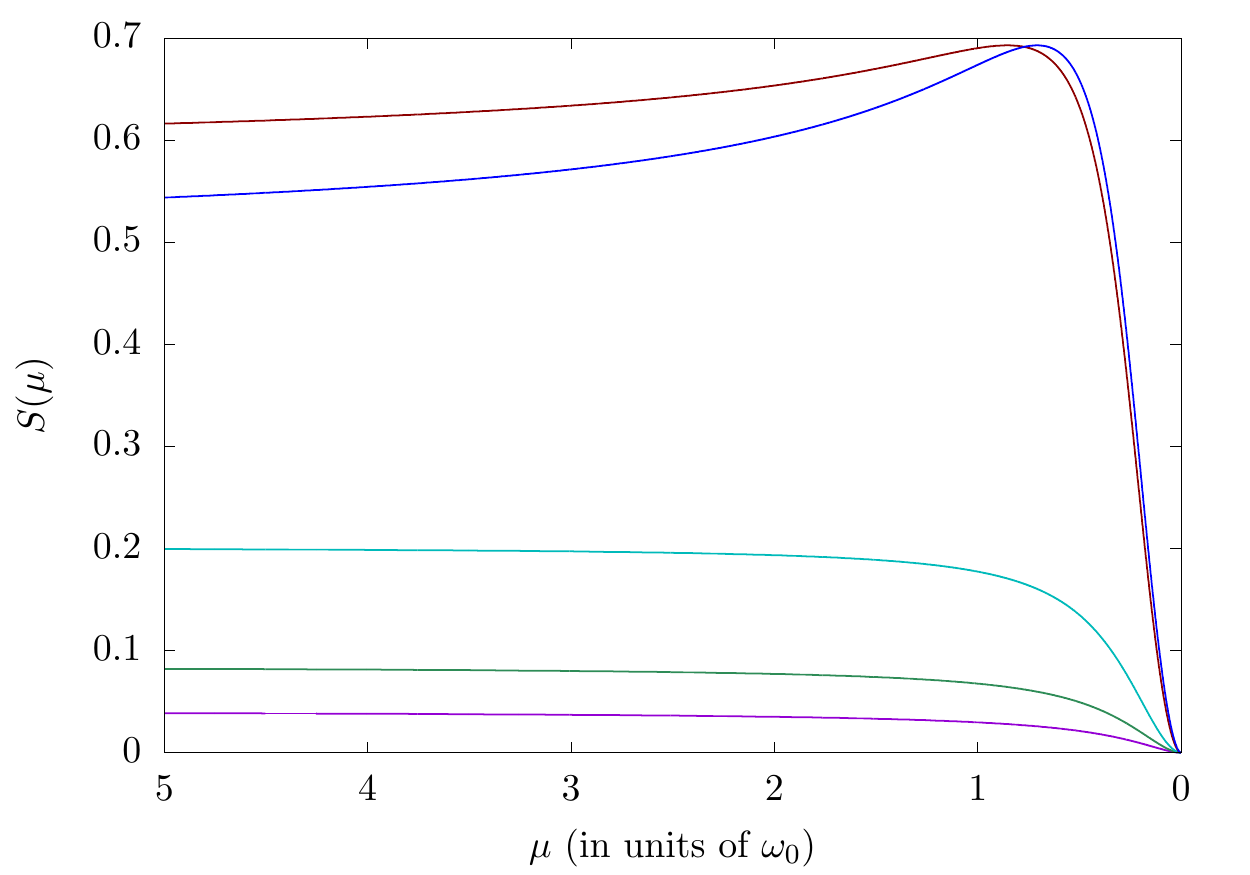}}
	~
	\subfloat[$m=1/2$\label{EnergyEnt2}]{
	\includegraphics[width=0.49\textwidth]{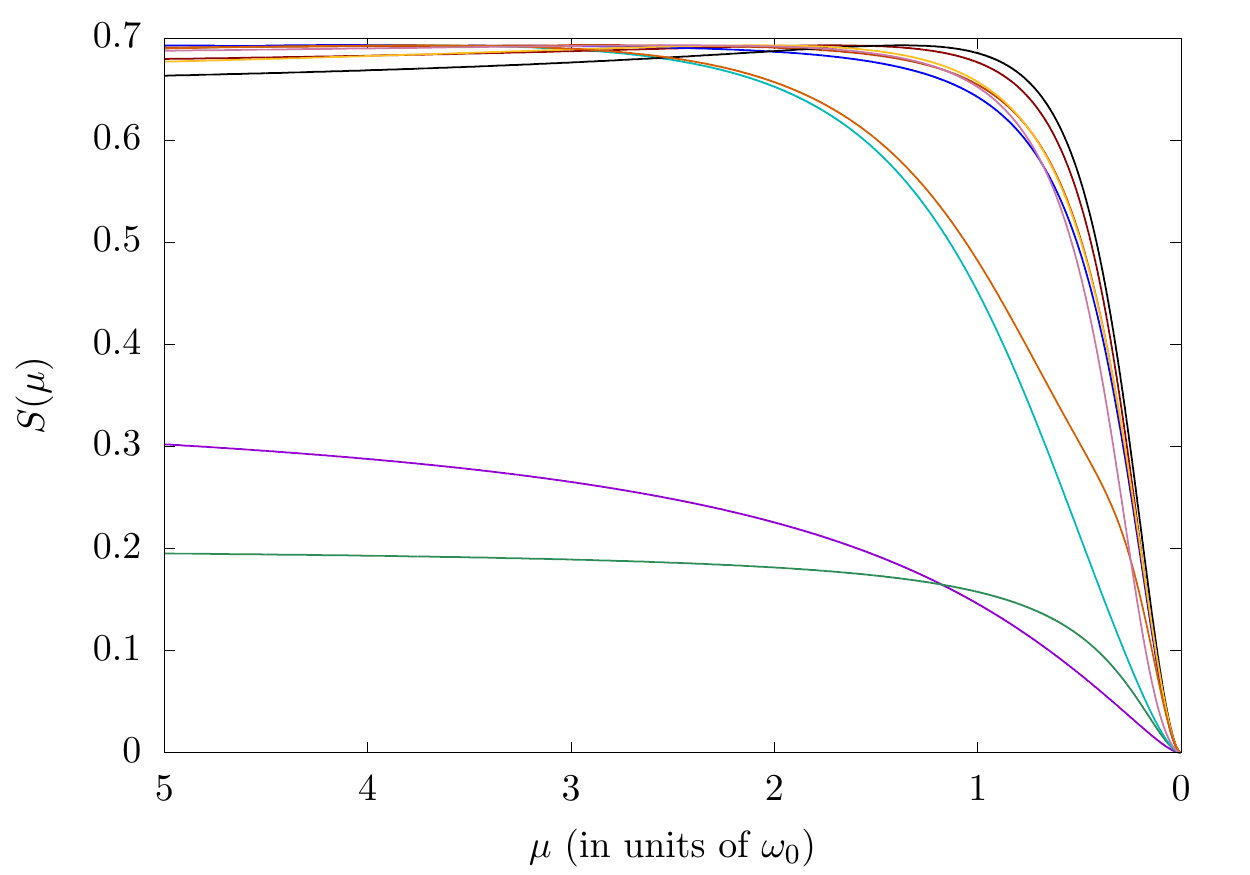}}

	\subfloat[$m=-3/2$\label{EnergyEnt3}]{
	\includegraphics[width=0.49\textwidth]{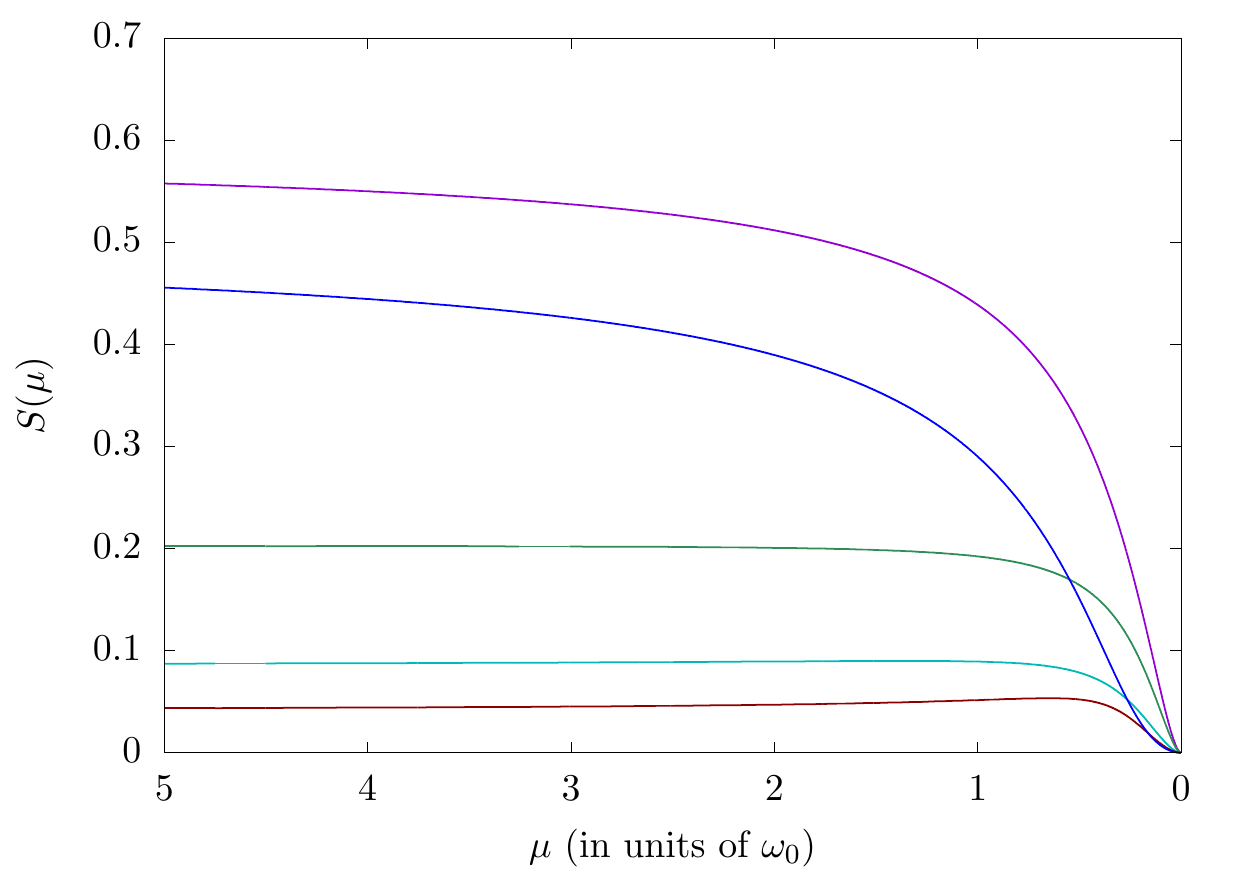}}
	~
	\subfloat[$m=-1/2$\label{EnergyEnt4}]{
	\includegraphics[width=0.49\textwidth]{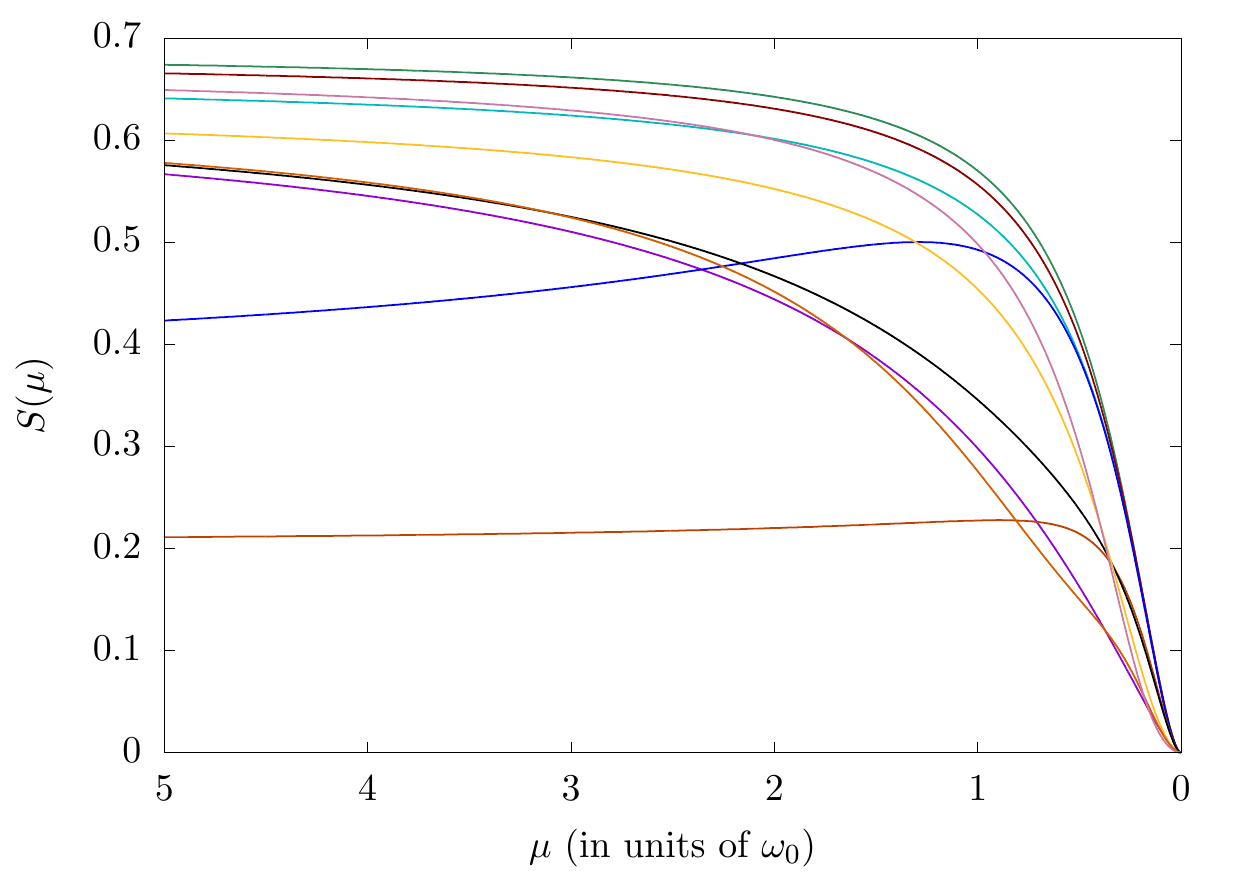}}
\end{center}
	\caption{
	The $\mu$ evolution of entanglement entropy $S(\mu)$ of all eigenstates of the Hamiltonian in Eq.~\eqref{singleH} with $N=5$. {We partition the system between the neutrino at $\omega_N$ and the remaining neutrinos.} Omitted are graphs of entanglement entropy for states with $m=\pm5/2$, as their corresponding $S(\mu)=0$ for all $\mu$. Here $m$ is the eigenvalue of $J^z$, i.e., half the difference between the numbers of $\nu_1$ and $\nu_2$ in the respective states. The number of solutions in each subfigure is $^N C_{N/2-m}$.}
	\label{EnergiesEntropy}
\end{figure*}

Using the methods outlined in Sec.~\ref{section:GenEntangle}, we may compute the entanglement entropy between any given neutrino and the rest of the ensemble, in any particular energy eigenstate $\ket{q}$, for generic values of $0<\mu<\mu_0$. First, we may express the state in either the mass or the flavor basis---here, we use the mass-basis density matrix $\rho(\mu)=V\ket{q}\!\bra{q}V^T$, and then reduce it by taking the partial trace over the remaining $N-1$ neutrinos.
For example, to obtain an entropy of entanglement between the neutrino with the highest frequency and the rest of the ensemble for a given energy eigenstate, we may use Eq.~\eqref{densityRed} for the reduced density matrix and Eq.~\eqref{EntEntropy} for the entropy. 
Results for energy eigenstates of a $N=5$ system are displayed in Fig.~\ref{EnergiesEntropy}. As one might expect, in the limit of $\mu \to 0$, {one obtains} $S \to 0$, as the energy eigenstates reduce simply to direct products of individual neutrino mass eigenstates. Notably, states with $m=\pm N/2$ (not shown in the figure) remain as direct products even for generic $\mu > 0$, and therefore continue to have $S=0$.

\section{Conditions for Adiabaticity} \label{Appendix_Ev}

It was noted in Refs.~\cite{Birol:2018qhx,Patwardhan:2019zta} that the Hamiltonian in Eq.~\eqref{singleH} has an energy spectrum exhibiting numerous level crossings. Some of these crossings are between energies of eigenstates with differing eigenvalues $m$ of the total $z$-component isospin $J^z$. 
Since $J^z$ is a time-independent conserved quantity of the Hamiltonian, one can show using Ehrenfest's theorem that the expectation value of $J^z$ with respect to any state is preserved as the state evolves through such a level crossing. Therefore, there can be no mixing between states with different eigenvalues $m$, thus rendering these crossings unimportant in the evolution of the system.

However, there are additional level crossings between energy eigenvalues of states with identical values of $m$. In this appendix, we 
\begin{enumerate}
    \item justify approximating our time evolution operator as $\mathcal{T}\{e^{-i\int_0^tH(t')dt'}\}\approx e^{-i\int_0^tH(t')dt'}$ from the analyticity of the Hamiltonian in time, and then 
    \item use the behavior of the commutators $[h_p(\mu),h_q(\mu')]$ to show that differing eigenvalues in $\vec{h}\equiv(h_1,\ldots,h_N)$ prevent mixing of degenerate states at the level crossings, including the states with the same $m$ values.
\end{enumerate}

It is helpful in demonstrating 
both claims listed above to first outline calculations for the commutators $[h_p(\mu),h_q(\mu')]$
. Using the definition of these charges in Eq.~\eqref{charges} and the fact 
\begin{align}
   \sum_{\substack{r=1\\r\neq p}}^N\sum_{\substack{s=1\\s\neq q}}^N\frac{[\vec{J}_p\cdot\vec{J}_r,\vec{J}_q\cdot\vec{J}_s]}{(\omega_p-\omega_r)(\omega_q-\omega_s)}=0,
\end{align}
it can be shown that our desired commutators are given by
\begin{align}
    [h_p(\mu),h_q(\mu')] &= 2\mu\sum_{\substack{r=1\\r\neq p}}\frac{[J_q^z,\vec{J}_p\cdot\vec{J}_r]}{\omega_p-\omega_r} -2\mu'\sum_{\substack{r=1\\r\neq q}}\frac{[J_p^z,\vec{J}_q\cdot\vec{J}_r]}{\omega_q-\omega_r}
    \nonumber \\
    &=\delta\mu\times
    \begin{cases} \sum_{\substack{r=1\\r\neq p}}^N\frac{J_p^+J_r^--J_p^-J_r^+}{\omega_p-\omega_r} & \mathrm{if}\: p=q \\ 
    \phantom{\sum_{r\neq p}}\mkern-18mu-\frac{J_p^+J_q^--J_p^-J_q^+}{\omega_p-\omega_q} & \mathrm{if}\: p\neq q {~.}\end{cases}
    \label{chargeComm}
\end{align}
Furthermore, using the identity
\begin{equation}
    H(\mu) = \sum_{p=1}^N\big(\omega_ph_p(\mu)+\mu \vec{J}_p\cdot\vec{J}_p\big)
    \label{chargesSum}
\end{equation}
it can also be shown that
\begin{align}
    [h_p(\mu),H(\mu')] &= \delta\mu \sum_{\substack{q=1\\q\neq p}}^N(J_p^+J_q^--J_p^-J_q^+) 
    \label{chargesCommHam} \\
    [H(\mu),H(\mu')] &= \delta\mu \sum_{p=1}^N\sum_{q=p+1}^N(\omega_p-\omega_q)(J_p^+J_q^--J_p^-J_q^+) {~.}
    \label{HamCommHam}
\end{align}
{\color{black} Notably, in the limit $\delta\mu\to0$, Eqs.~\eqref{chargeComm} and \eqref{chargesCommHam} reduce to the well-known commutation relations between the conserved charges.}


Recall that a Magnus expansion of the Dyson series is given by
\begin{align}
    \mathcal{T}\bigg\{e^{-i\int_{t}^{t+\delta t}H(t')dt'}\bigg\} = \exp\bigg[-i\int_{t}^{t+\delta t}H(t')dt' \nonumber \\
    +\frac{i}{2}\int_{t}^{t+\delta t}\int_t^{t'}[H(t'),H(t'')]dt''dt'+\cdots\bigg]
\end{align}
for $\int_{t}^{t+\delta t}||H(t')||_2dt' <\pi$ with the matrix norm $||X||_2\equiv\big[\sum_{i,j}|X_{ij}|^2\big]^{1/2}$ \cite{MagnusBound}.
Since our Hamiltonian is analytic in time, we can use the fact that $[H(t),H(t')] \sim \mathcal{O}(\delta t)$, for sufficiently small time steps $\delta t\equiv t'-t$, to approximate this expansion by the lowest order term in $\delta t$---as all terms with commutators will be $\lesssim\mathcal{O}[(\delta t)^3]$. This commutator relationship is also corroborated by Eq.~\eqref{HamCommHam}. Combining these approximate terms $\mathcal{T}\left\{e^{-i\int_t^{t+\delta t}H(t')dt'}\right\}\approx e^{-i\int_t^{t+\delta t}H(t')dt'+\mathcal{O}[(\delta t)^3]}$ for all time steps between $0$ and $t$, we obtain 
\begin{equation}
\mathcal{T}\left\{e^{-i\int_0^{t}H(t')dt'}\right\}\approx \exp\left[-i\int_0^{t}H(t')dt'+N_t\,\mathcal{O}[(\delta t)^3]\right],
\end{equation}
where $N_t \sim t/\delta t$. This argument can similarly be applied to taking small steps in $\delta\mu$, as these steps can be related to small time steps, given a function $\mu(t)$ such as Eq.~\eqref{singleMu} with a well-behaved derivative, $d\mu/dt$.\footnote{While our form for $d\mu/dt$ implies that $\delta t$ will grow as $\mu\to0$ if step sizes $\delta\mu$ are held fixed, in this limit we find that the nonlinear term of our Hamiltonian in Eq.~\eqref{singleH} {becomes negligible}, in which case our Hamiltonian would commute with itself at varying times and make our approximate time evolution operator exact regardless.} Thus, we propose the approximation of the time evolution operator used in Eq.~\eqref{basicTimeEvolve}, justifying claim 1 stated earlier.

Next we turn our attention to claim 2. Consider an energy eigenstate with charge eigenvalues $\vec{\epsilon}(t)$ for times $0\leq t\leq t_d$, where $t_d$ is a time at which this state shares the same energy eigenvalue with another eigenstate of $\vec{h}$. The time evolution of this state to values beyond $t_d$ over a small time interval $(t_d-\delta,t_d+\delta)$ can be described by $\exp\big[-i\int_{t_d-\delta}^{t_d+\delta}H(t')dt'\big]\ket{\vec{\epsilon}(t_d-\delta)}$; moreover, applying the operator $h_p$ to this state yields
\begin{align}
    h_p(&t_d+\delta)\exp\bigg[-i\int_{t_d-\delta}^{t_d+\delta}H(t')dt'\bigg]\ket{\vec{\epsilon}(t_d-\delta)} 
    \nonumber \\
    &\approx \epsilon_p(t_d-\delta)\exp\bigg[-i\int_{t_d-\delta}^{t_d+\delta}H(t')dt'\bigg]\ket{\vec{\epsilon}(t_d-\delta)} 
    \nonumber \\
    & \phantom{\approx}+\bigg[h_p(t_d+\delta),\exp\bigg[-i\int_{t_d-\delta}^{t_d+\delta}H(t')dt'\bigg]\bigg]\ket{\vec{\epsilon}(t_d-\delta)} 
    \label{adiabQuery}
\end{align}
where we use the approximation $h_p(t_d+\delta)\ket{\vec{\epsilon}(t_d-\delta)}\approx \epsilon_p(t_d-\delta)\ket{\vec{\epsilon}(t_d-\delta)}+\mathcal{O}(\delta)$. 

Furthermore, we can observe for the latter term of Eq.~\eqref{adiabQuery} that 
\begin{align}
    \bigg[h_p&(t_d+\delta),\exp\bigg[-i\int_{t_d-\delta}^{t_d+\delta}H(t')dt'\bigg] \bigg]
    \nonumber \\
    =&-i\int_0^1\exp\bigg[-i(1-s)\int_{t_d-\delta}^{t_d+\delta}H(t')dt'\bigg] 
    \nonumber \\
    &\times \int_{t_d-\delta}^{t_d+\delta}[h_p(t_d+\delta),H(t')]dt'
    \nonumber \\
    &\times \exp\bigg[-is\int_{t_d-\delta}^{t_d+\delta}H(t')dt'\bigg]ds {~.}
\end{align} 
Using our earlier calculation in Eq.~\eqref{chargesCommHam}, we can see that this term is $\mathcal{O}(\delta^2)$ and therefore may be neglected for $\delta \to 0$. Taking this limit, we thus explicitly see that states with different eigenvalues $\vec{\epsilon}$ of $\vec{h}$ are prevented from mixing with one another as they evolve through the crossing.

\bibliography{references}

\end{document}